\documentclass[10pt,aps,pre,twocolumn,superscriptaddress,showpacs,floatfix]{revtex4-2}
\usepackage{amsmath}
\usepackage{amssymb}
\usepackage{amsfonts}
\usepackage{lipsum}

\usepackage[ruled,vlined]{algorithm2e}

\usepackage{graphicx}
\usepackage[normalem]{ulem}

\usepackage{bm}
\usepackage{braket}
\usepackage{mathtools}

\usepackage{hyperref}
\hypersetup{colorlinks=true,linkcolor=blue,citecolor=blue,urlcolor=blue}

\begin{document}

\title{Quantum-echo Markov process for combinatorial optimization}

\author{Tatsuhiko Shirai}
\email{tatsuhiko.shirai@aoni.waseda.jp}
\affiliation{Waseda Institute for Advanced Study, Waseda University, Nishi Waseda, Shinjuku-ku, Tokyo 169-0051, Japan}

\date{\today}

\begin{abstract}
    We introduce a quantum-echo Markov process for combinatorial optimization.
    Quantum dynamics based on quantum annealing (QA) or the quantum approximate optimization algorithm (QAOA) is used to engineer the transition kernel.
    Increasing the annealing time in QA or the number of layers in QAOA enables transitions to explore distant configurations while suppressing large energy changes.
    The Hamming-space delocalization originates from operator spreading, whereas the energy-space localization arises from a dynamically generated correlation.
    For optimization, we propose quantum-echo local optimization.
    For the ferromagnetic Ising chain, increasing the annealing time or circuit depth yields a nearly quadratic improvement in the dependence of the mean first-passage time on the inverse optimality gap compared with random single-spin flips.
    For the random Ising model, increasing these parameters initially improves optimization performance, whereas further increases degrade performance in QA and lead to saturation in QAOA.
    Quantum-echo greedy descent substantially improves the performance, highlighting the complementary roles of quantum dynamics in exploration and greedy descent in exploitation.
    These results establish quantum-echo dynamics as a framework for engineering structured transition kernels and provide a route to using finite-resource quantum many-body dynamics as a computational primitive for iterative optimization.
\end{abstract}

\maketitle

\section{Introduction}
Recent advances in quantum hardware have enabled controlled realizations of quantum many-body dynamics across a variety of programmable quantum platforms~\cite{barrends2016digitized,zhang2017observation,ebadi2021quantum,king2023quantum,miessen2024benchmarking,manovitz2025quantum,abanin2025observation}.
A fundamental question is whether such dynamics can be used as a computational resource to address practically important problems.
One promising application is combinatorial optimization~\cite{lucas2014ising}, which aims to minimize a cost function on an exponentially large set of candidate configurations.
Combinatorial optimization arises in many real-world applications, including logistics, drug discovery, and scheduling.

Quantum annealing (QA)~\cite{kadowaki1998quantum,chakrabarti2023quantum} and the quantum approximate optimization algorithm (QAOA)~\cite{farhi2014quantum, blekos2024review} have attracted considerable attention as quantum approaches to combinatorial optimization.
Throughout this work, we focus on the finite-resource regime in which the annealing time in QA and the number of alternating layers in QAOA remain independent of the problem size.
Recent studies have revealed limitations of QA and QAOA in this regime.
For optimization problems on sparse graphs, finite-time QA and fixed-depth QAOA~\footnote{Here, the depth refers to the number of alternating QAOA layers rather than the compiled circuit depth.} exhibit an effective locality structure: the influence of their dynamics is restricted to bounded neighborhoods of the underlying graph~\cite{hastings2019classical,farhi2020quantum2,farhi2020quantum1,moosavian2022limits}.
Consequently, for certain problems and graph families, these algorithms cannot match the approximation performance of known classical algorithms, including classical local algorithms~\cite{bravi2020obstacles,farhi2020quantum1,farhi2020quantum2,barak2022classical,moosavian2022limits}.
These results apply to specific algorithmic settings and do not preclude improved quantum performance in other problem classes and instance families~\cite{farhi2022quantum,basso2022performance,apte2026quantum}, at annealing times or QAOA depths beyond the regimes covered by the corresponding bounds~\cite{somma2012quantum,farhi2020quantum1,farhi2020quantum2,hastings2021power}, in open-system QA implementations~\cite{bauza2025scaling}, or when quantum dynamics is embedded in broader hybrid~\cite{shirai2024postprocessing} or fault-tolerant quantum algorithms~\cite{ruslan2024evidence}.
Nevertheless, they motivate us to explore a different use of finite-time QA and fixed-depth QAOA, not as standalone optimizers that produce a final solution.

\begin{figure}[t]
    \centering
    \includegraphics[width=0.96\linewidth]{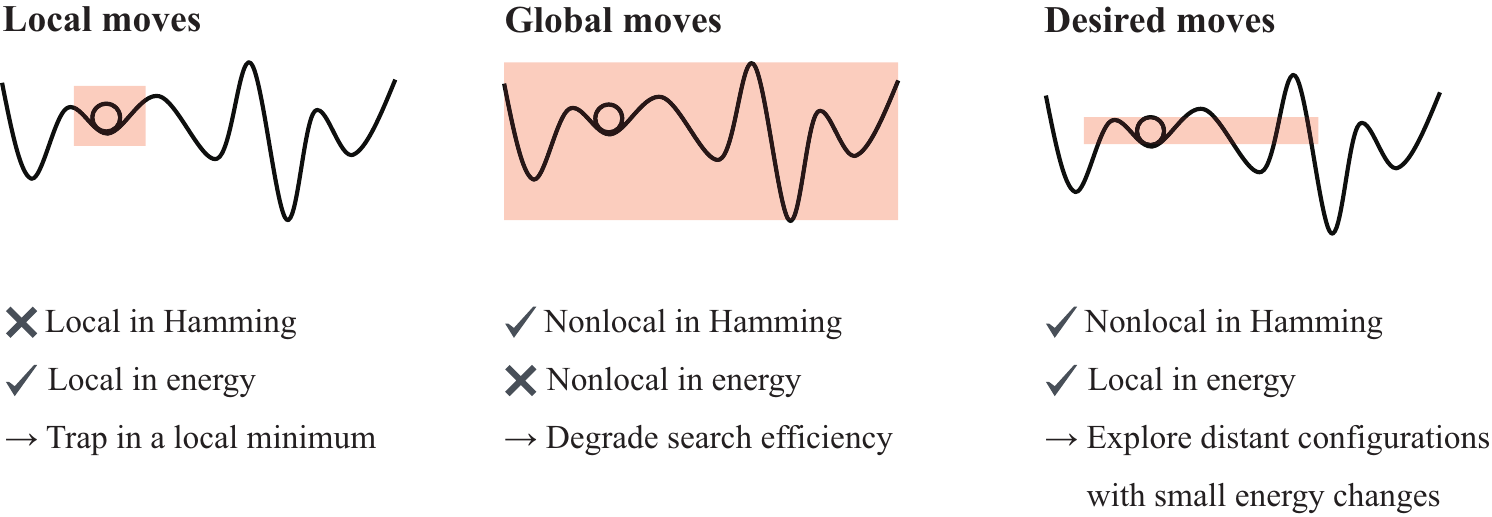}
    \caption{
    Schematic illustration of three types of moves generated by a transition matrix.
    The circle denotes the current configuration, and the shaded region represents the candidate configurations.
    Hamming-local moves can be trapped in a local minimum, whereas moves that are nonlocal in both Hamming and energy spaces can degrade search efficiency.
    The desired transitions are nonlocal in Hamming space but local in energy space, allowing the search to explore distant configurations with small energy changes.
    }
    \label{fig:schematic}
\end{figure}

We use QA and QAOA as quantum primitives for constructing nontrivial moves within an iterative search process.
This process can be described by a transition matrix that specifies the  probability of moving from the current configuration to candidate configurations~\cite{kampen2007stochastic}.
As illustrated in Fig.~\ref{fig:schematic}, moves that are too local in Hamming space can fail to escape local minima, whereas unstructured nonlocal moves typically involve large changes in cost and can degrade search efficiency.
We refer to the latter property as nonlocality in energy space.
It is therefore desirable to generate transitions that are nonlocal in Hamming space but still local in energy space.
Such transitions allow the search to explore distant configurations while keeping energy changes small.

Quantum dynamics has recently been employed to construct proposal distributions for Markov chain Monte Carlo (MCMC).
In quantum-enhanced MCMC, unitary quantum dynamics generates proposals that can connect configurations separated in Hamming space while maintaining relatively small energy differences and has been shown to accelerate Boltzmann sampling~\cite{layden2023quantum}.
Subsequent studies have shown that efficient sampling requires an intermediate regime in which quantum dynamics sufficiently delocalizes the state over classical configurations without generating excessive entropy~\cite{orfi2024barriers}.
Theoretical limitations for unstructured sampling problems have also been established~\cite{orfi2024bounding}.
Furthermore, time-dependent protocols based on alternating quantum circuits or QA have been proposed to improve sampling efficiency~\cite{nakano2024markov,christmann2025from,arai2025quantum,hsieh2025adiabatic} or optimization performance~\cite{zhu2026combinatorial}.

Motivated by recent developments in quantum-enhanced MCMC, we employ a quantum-echo construction to generate transition probabilities.
In the absence of unitary dynamics, the resulting transitions are local single-spin flips, allowing us to investigate how quantum dynamics modifies the locality of transitions in Hamming and energy spaces.
By a quantum-echo process, we mean forward evolution, a local perturbation, and backward evolution, represented by $\hat{U} \hat{V}_i \hat{U}^\dagger$~\cite{pappalardi2020quantum}. 
Closely related forward-perturbation-backward protocols have been used to investigate operator spreading through out-of-time-order correlators (OTOCs)~\cite{larkin1969quasiclassical,garttner2017measuring,li2017measuring,braumuller2022probing,abanin2025observation}.
More broadly, the combination of many-body evolution and local operations employed in our QA and QAOA-based echo sequences is naturally compatible with digital-analog quantum computing architectures~\cite{parrarodriguez2020digital,headley2022approximating,bluvstein2022quantum,andersen2025thermalization,deshpande2026analog-digital}.
Unlike conventional digital-analog protocols designed to realize target unitary dynamics, here the echo-dressed local operations are used to construct a Markov transition kernel with tunable locality.
We term the resulting process the quantum-echo Markov process.

Applying this framework to the random Ising model and the random energy model (REM)~\cite{derrida1981random}, we show that increasing the annealing time in QA or the number of QAOA layers simultaneously delocalizes transitions in Hamming space and localizes them in energy space.
Remarkably, the resulting transitions can be more local in energy space than single-spin flips, despite being more nonlocal in Hamming space.
We further elucidate the underlying mechanisms.
The Hamming-space delocalization originates from operator spreading of $\hat{V}_i$ under the unitary dynamics, while the energy-space localization is associated with a correlation between the cost and the Hamming weight of the eigenstates of the effective Hamiltonian $\hat{C}_\mathrm{eff}=\hat{U}^\dagger\hat{C} \hat{U}$, where $\hat{C}$ is the problem Hamiltonian.
The emergence of this correlation may be related to the decrease in $\bra{+} \hat{C}_\mathrm{eff} \ket{+}$, where $\ket{+}$ denotes the uniform superposition of all computational-basis states, with increasing annealing time in QA or number of layers in QAOA.
Furthermore, our numerical results for the random Ising model suggest that the energy-space localization parameter remains nearly independent of system size.

For optimization, we first propose quantum-echo local optimization, in which transitions generated by the quantum-echo Markov process are accepted only when they do not increase the cost function.
For the ferromagnetic Ising chain, we identify a crossover between a thermal-state-like regime at high energies and a diffusion-annihilation regime at low energies.
The two regimes exhibit distinct scaling behaviors of the iteration count with the optimality gap.
As the annealing time or the number of layers increases, the thermal-state-like regime extends toward smaller optimality gaps, providing a nearly quadratic speedup in the gap dependence of the mean first-passage time compared with random single spin flips.
We interpret this crossover in terms of the number of local defects.
For the random Ising model, increasing these parameters initially improves optimization performance, whereas further increases degrade performance in QA and lead to saturation in QAOA.
We then consider an enhanced strategy, quantum-echo greedy descent, in which each quantum-echo transition is followed by a greedy single-spin-flip descent.
Adding greedy descent significantly improves the optimization performance, even in regimes where strong energy-space localization limits the performance of quantum-echo local optimization alone.
This improvement highlights the complementary roles of quantum and classical dynamics: quantum dynamics provides exploration, whereas greedy descent provides exploitation.
These results demonstrate that quantum many-body dynamics can be systematically incorporated into a Markov process to improve approximate optimization.

The rest of this paper is organized as follows.
Section~\ref{sec:methods} introduces the general framework of the quantum-echo Markov process.
Section~\ref{sec:model} describes the methods and the models used in this work.
Section~\ref{sec:results} demonstrates how locality in the Hamming and energy spaces can be tuned and elucidates the underlying mechanism.
Section~\ref{sec:application} introduces quantum-echo local optimization and quantum-echo greedy descent and presents their applications.
Section~\ref{sec:conclusion} concludes the paper and discusses future directions.

\section{Quantum-echo Markov process}~\label{sec:methods}
We adopt a binary encoding of combinatorial optimization problems, where the cost function is represented by $C_z$ with $z=\{z_i\}_{i=1}^N$.
Here, $z_i \in \{0, 1\}$ denotes a spin variable and $N$ is the number of spins.
Many combinatorial optimization problems, such as Max-Cut, can be represented by this form~\cite{lucas2014ising}.
In this work, we assume that an optimum solution corresponds to a spin configuration that minimizes $C_z$.

The Markov process iteratively updates the configuration based only on its current state.
Then, a transition matrix $T_{zz'}$ determines the probability of moving from the current configuration $z$ to a candidate configuration $z'$.

To define the quantum-echo Markov process, we consider a unitary operator $\hat{U}$ and a set of local unitary operators $\{\hat{V}_i\}_{i=1}^N$.
The state after the quantum-echo dynamics is given by
\begin{equation}
    \ket{\psi_i(z)}=\hat{U} \hat{V}_i \hat{U}^\dagger \ket{z}.
    \label{eq:quantu_echo}
\end{equation}
Here, $\ket{z}$ is a computational-basis state.
The specific choices of $\hat{U}$ and $\{\hat{V}_i\}$ used in this work are described in Sec.~\ref{sec:model}.
Then, the transition matrix is represented as
\begin{equation}
    T_{zz'} =\frac1{N}\sum_{i=1}^N |\langle z'\ket{\psi_i(z)}|^2=\frac{1}{N}\sum_{i=1}^N |\bra{z_U'} \hat{V}_i \ket{z_U}|^2,
    \label{eq:transition_probability}
\end{equation}
where $\ket{z_U}=\hat{U}^\dagger \ket{z}$.
The uniform selection of $\hat{V}_i$ and the quantum measurement of $\ket{\psi_i(z)}$ in the computational basis can sample the candidate configuration according to the transition probability.
When $\hat{V}_i$ is Hermitian, the transition probability is symmetric with respect to $z$ and $z'$.
Note that $T_{zz'}$ satisfies the necessary properties of the transition matrix: $T_{zz'} \geq 0$ and $\sum_{z'} T_{zz'}=1$.

Then, we introduce locality in Hamming space and energy space.
The locality in Hamming space uses the Hamming distance $D_{zz'}=\sum_{i=1}^N |z_i -z'_i|$, which counts the number of flipped spins, and is defined as 
\begin{align}
    d_H  = 2^{-N}\sum_{z,z'} D_{zz'} T_{zz'}.
\end{align}
On the other hand, the locality in energy space is defined as
\begin{align}
    d_E = 2^{-N}\sum_{z,z'} \Delta C_{zz'}^2 T_{zz'},
\end{align}
which measures the typical squared energy change induced by the transitions.
Here, $\Delta C_{zz'}=C_{z'}-C_z$.

\section{Methods and models}\label{sec:model}
\begin{figure*}[t]
    \centering
    \includegraphics[width=0.48\linewidth]{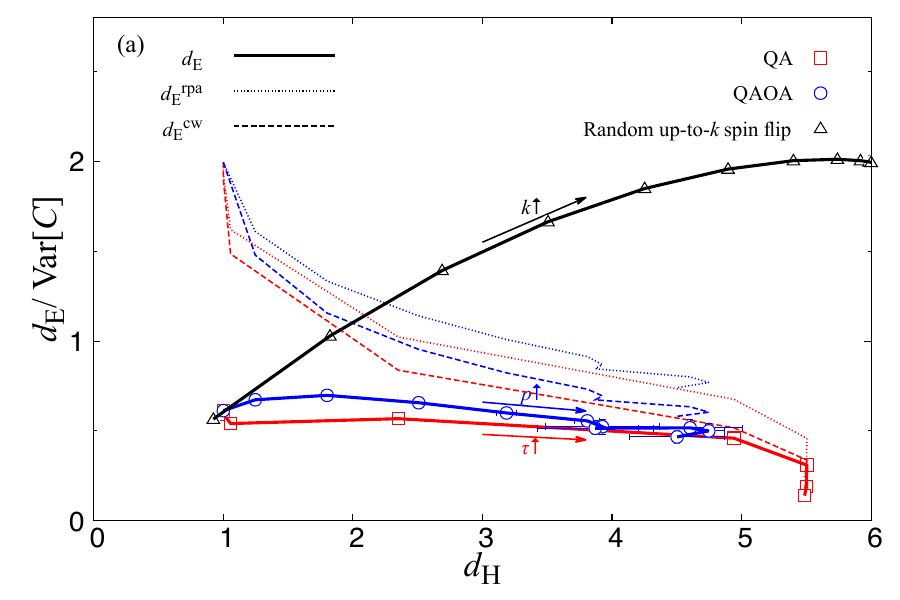}
    \includegraphics[width=0.48\linewidth]{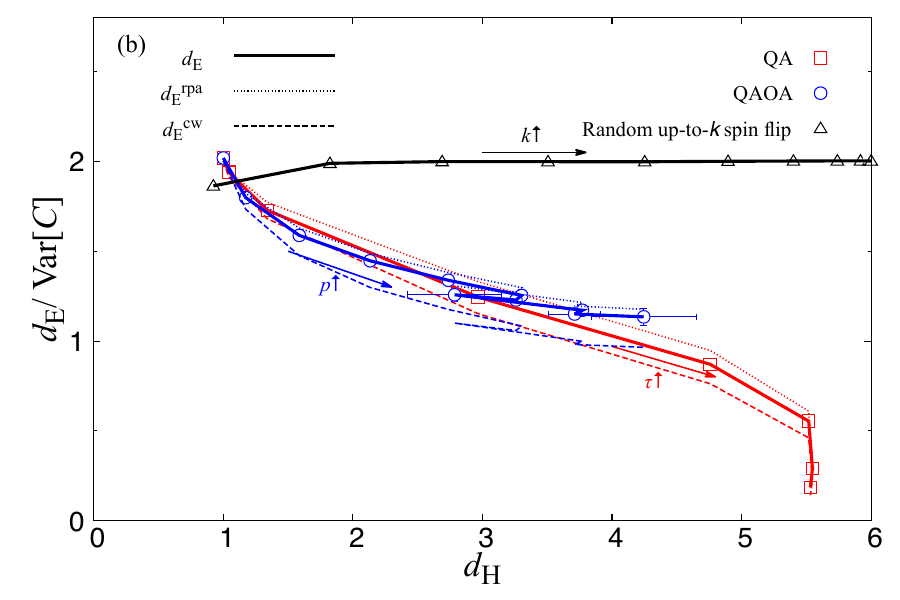}
    \caption{
    Energy-space locality $d_E$ versus Hamming-space locality $d_H$ for (a) the random Ising model and (b) the REM.
    Solid lines show the exact $d_E$, whereas dotted lines and dashed lines represent approximations $d_E^\mathrm{rpa}$ and $d_E^\mathrm{cw}$, respectively.
    Squares (red) and circles (blue) denote the quantum-echo Markov process based on QA and QAOA, respectively, and triangles (black) denotes the random up-to-$k$ spin flips.
    Arrows indicate increasing $p$, $\tau$, and $k$.
    In the REM, both approximations closely reproduce the exact result, whereas in the random Ising model, a deviation between $d_E$ and $d_E^\mathrm{rpa}$ appears for small $\tau$ and $p$.
    We set $N=12$ for both models.
    }
    \label{fig:dE_dH}
\end{figure*}

\subsection{Quantum dynamics}
In this work, we implement the unitary dynamics of the quantum-echo Markov process using either QA or QAOA.
For the local operation, we choose $\hat{V}_i=\hat{\sigma}_i^x$, where $\hat{\bm \sigma}_i= (\hat{\sigma}_i^x, \hat{\sigma}_i^y, \hat{\sigma}_i^z)$ denotes the Pauli operators that act on the $i$-th spin.
With this choice, the quantum-echo Markov process reduces to a random single-spin-flip process in the absence of unitary dynamics.
In our numerical calculations, the quantum-echo Markov process is simulated on classical computers, while its implementation on quantum hardware is left for future work.

\subsubsection{Quantum annealing}
In QA, the unitary operator in Eq.~(\ref{eq:quantu_echo}) is provided as
\begin{equation}
    \hat{U} = \mathcal{T} \exp \left( -\mathrm{i}\int_0^\tau \hat{H} \left(\frac{t}{\tau} \right) dt\right),
\end{equation}
where $\mathcal{T}$ is the time-ordering operator and $\tau$ is the annealing time.
The dynamics involves the transverse-field and cost Hamiltonians,
\begin{equation}
    \hat{H}_X = -\sum_{i=1}^N \hat{\sigma}_i^x, \quad    \hat{H}_C = \frac{\hat{C}}{\mathcal{N}},
    \label{eq:two_Hamiltonians}
\end{equation}
where $\hat{C}\ket{z}=C_z\ket{z}$ and $\mathcal{N}$ is a normalization factor.
The time dependent Hamiltonian linearly interpolates between $\hat{H}_X$ and $\hat{H}_C$: for $s \in [0, 1]$
\begin{equation}
    \hat{H}(s) = s \hat{H}_C +(1-s) \hat{H}_X.
\end{equation}
We consider annealing times $\tau\in \{0, 1, 2, 5, 10, 20, 50, 100\}$.

\subsubsection{Quantum approximate optimization algorithm}
In QAOA, the unitary operator in Eq.~(\ref{eq:quantu_echo}) is given as
\begin{equation}
    \hat{U} = \hat{U}_X(\beta_p) \hat{U}_C(\gamma_p) \cdots \hat{U}_X(\beta_1) \hat{U}_C(\gamma_1),
\end{equation}
where $\hat{U}_X(\beta)=\exp(-\mathrm{i} \beta \hat{H}_X)$, $\hat{U}_C(\gamma)=\exp(-\mathrm{i} \gamma \hat{H}_C)$, and $p$ denotes the number of layers.
The variational parameters $\{ \beta_i\}_{i=1}^p$ and $\{ \gamma_i\}_{i=1}^p$ are optimized by minimizing the energy expectation value $\bra{+} \hat{C}_\mathrm{eff} \ket{+}$.

For implementation, the number of layers is chosen from $p \in \{0,1,\ldots, 10\}$.
For each number of layers and each problem instance, the variational parameters are optimized using the Powell method implemented in SciPy.
Ten independent optimization runs are performed with different initial sets $(\{\beta_i\}_{i=1}^p, \{\gamma_i\}_{i=1}^p)$.
Each initial set is generated by sampling $\{\beta_i\}_{i=1}^p$ and $\{\gamma_i\}_{i=1}^p$ independently from the interval $[-\pi/(2N),\pi/(2N)]$ following Ref.~\cite{park2024hamiltonian}.
We set ${\it ftol}=0.001$ and use the default values for all other options.
Among the ten optimization runs, we select the parameter set that minimizes $\bra{+} \hat{C}_\mathrm{eff} \ket{+}$ excluding any parameter set for which the resulting transition matrix becomes the identity matrix, as described in Appendix~\ref{appendix:SC_RIM}.

\subsection{Random up-to-$k$-spin flips}
As a baseline, we choose the transition probability to be uniform over all configurations that satisfy $D_{zz'} \leq k$, and zero otherwise.
The transition probability is normalized to satisfy $\sum_{z'} T_{zz'}=1$.
We refer to this process as random up-to-$k$-spin flips.
This generates Hamming-space local moves for $k=O(1)$ and allows Hamming-space nonlocal moves for $k=\Theta(N)$.
For fixed $k$ and large $N$, almost all candidate configurations lie at Hamming distance $k$, so the process asymptotically reduces to random $k$-spin flips.

\subsection{Models}
We consider two representative models with distinct energy-landscape structures: the random Ising model and the REM.

In the random Ising model, the cost function is given as
\begin{equation}
    C_z= -\sum_{i=1}^N \sum_{j=i+1}^N J_{ij} (-1)^{z_i+ z_j} - \sum_{i=1}^N h_i (-1)^{z_i},
\end{equation}
where $J_{ij}$ and $h_i$ are independently sampled from the probability distributions,
\begin{align}
    \rho(J_{ij})=&\frac{1}{2} \left[\delta (J_{ij}-1)+\delta (J_{ij}+1)\right],\nonumber\\
    \rho(h_{i})=&\frac{1}{2} \left[ \delta (h_{i}-1)+\delta (h_{i}+1)\right].
\end{align}
Because the cost function is $2$-local, the energies of the configurations are correlated with the Hamming-space structure.

In the REM, the cost function is independently sampled from a Gaussian distribution,
\begin{equation}
    \rho(C_z)= \frac{1}{\sqrt{2\pi N}} \exp \left( -\frac{C_z^2}{2 N} \right).
\end{equation}
In contrast to the random Ising model, the REM has no intrinsic correlation between energy and Hamming-space structure.
Here, the REM is used as an idealized structureless benchmark, assuming oracle access to the unitary evolution generated by its cost Hamiltonian in both QA and QAOA.

For each model and system size $N$, we generate five random instances and evaluate the mean and standard deviation over these instances.
For both models, we set the normalization in Eq.~(\ref{eq:two_Hamiltonians}) to $\mathcal{N}=\sqrt{N}$ so that a single-spin flip typically changes the normalized cost by $O(1)$.

\section{Locality generation and mechanism}\label{sec:results}
Figures~\ref{fig:dE_dH}~(a) and (b) show $d_H$ versus $d_E$ for quantum-echo Markov processes based on QA and QAOA and for random up-to-$k$-spin flips.
Random up-to-$k$-spin flips do not simultaneously achieve Hamming-space nonlocality and energy-space locality.
In contrast, the quantum-echo Markov process exhibits increasing Hamming-space nonlocality while maintaining energy-space locality as the annealing time or number of layers increases.
Despite this common tendency, the two models exhibit different behaviors in energy space.
In the random Ising model, $d_E$ is nearly independent of $p$ and $\tau$, whereas in the REM, $d_E$ decreases with $p$ and $\tau$.
In the following, we explain the mechanisms of delocalization in Hamming space and localization in energy space, as well as the origin of the difference between the two models.

To explain the locality in Hamming space, we rewrite $d_H$ as squared commutators, or equivalently, the OTOC expression:
\begin{equation}
    d_H = \frac{1}{4N} \sum_{i,j}^N\biggl\|  [\hat{U}^\dagger \hat{\sigma}_j^z \hat{U}, \hat{\sigma}_i^x] \biggr\|_\mathrm{F}^2,
    \label{eq:locality_Hamming}
\end{equation}
where $\| \hat{O} \|_\mathrm{F} = \sqrt{2^{-N}\operatorname{Tr} (\hat{O}^\dagger \hat{O})}$ is the scaled Frobenius norm.
This expression shows that $d_H$ increases as $\hat{U}^\dagger \hat{\sigma}_j^z \hat{U}$ becomes nonlocal in operator space.
Here, nonlocality means the growth of operator size~\cite{bentsen2019fast}, namely the emergence of high-order Pauli strings involving many spins.
As the annealing time or the number of layers increases, low-order Pauli strings are converted to high-order Pauli strings, leading to delocalization in Hamming space.

To clarify the origin of localization in energy space, we introduce $d_E^\mathrm{rpa}$ and $d_E^\mathrm{cw}$ (see Appendix~\ref{appendix:derivation} for their derivations).
Under a random-phase approximation, $d_E$ is approximated by
\begin{equation}
    d_E^\mathrm{rpa}=2^{-N}\sum_{z,z'} \Delta C_{zz'}^2 T_{zz'}^\mathrm{rpa},
\end{equation}
where the transition probability $T_{zz'}$ is replaced by $T_{zz'}^\mathrm{rpa}=\sum_x p_x^z p_x^{z'}$, with $p_x^z=|\langle x \ket{z_U}|^2$ and $\hat{\sigma}_i^x\ket{x} =(-1)^{x_i}\ket{x}$.
Throughout this work, $\ket{x}$ and $\ket{z}$ denote the eigenstate of $\hat{\sigma}_i^x$ and $\hat{\sigma}_i^z$, respectively.
This approximation neglects the correlations among the phases of the amplitude $\langle x \ket{z_U}$.

The dotted lines in Figs.~\ref{fig:dE_dH} show the $\tau$ and $p$ dependence of $d_E^\mathrm{rpa}$.
For the random Ising model, $d_E^\mathrm{rpa}$ decreases with $\tau$ or $p$ and approaches $d_E$ as $\tau$ or $p$ increases, indicating that this approximation becomes more accurate at larger $\tau$ and $p$.
In contrast, for the REM, $d_E^\mathrm{rpa}$ provides a good approximation to $d_E$ throughout the range of $\tau$ and $p$ considered.

To clarify the difference between the random Ising model and the REM, we first consider the case $p=0$ or $\tau=0$, where $\hat{U}$ is the identity matrix, for which the exact transition corresponds to random single-spin flips.
In this case, $T_{zz'}^\mathrm{rpa}=2^{-N}$, and hence $T_{zz'}^\mathrm{rpa}$ is completely delocalized in Hamming space, whereas $T_{zz'}$ is restricted to the Hamming space one.
For the random Ising model, a single-spin flip yields $d_E/\operatorname{Var}[C]=O(N^{-1})$, whereas the random-phase approximation gives $d_E^\mathrm{rpa}/\operatorname{Var}[C]=O(N^0)$, where $\operatorname{Var}[C]=\overline{C_z^2} -(\overline{C_z})^2$ and the overline denotes the average over computational-basis states.
Thus, the large discrepancy at small $\tau$ and $p$ originates from the fact that the random-phase approximation replaces Hamming-local transitions by Hamming-nonlocal transitions.
As $\tau$ or $p$ increases and the exact transition becomes more delocalized in Hamming space, this replacement becomes less severe.
For the REM, by contrast, since the energy difference is independent of the Hamming distance, replacing Hamming-local transitions by Hamming-nonlocal ones does not change the characteristic energy difference.
Consequently, both $d_E/\operatorname{Var}[C]$ and $d_E^\mathrm{rpa}/\operatorname{Var}[C]$ are $O(N^0)$.

Despite this difference in the accuracy of the random-phase approximation, $d_E^\mathrm{rpa}$ decreases with $\tau$ and $p$ for both the random Ising model and the REM.
The dashed lines in Figs.~\ref{fig:dE_dH} show that this dependence is well captured by
\begin{equation}
    d_E^\mathrm{cw}=2^{-2N} A \sum_{z,z'} \Delta C_{zz'}^2 \exp \left(- \frac{\Delta C_{zz'}^2}{\xi \operatorname{Var}[C]}\right),
\end{equation}
where
\begin{align}
    A=&\left[ 1 -\left( \frac{4 \operatorname{Var}[w]}{N} \right)^2 \right]^{-1/2}, \nonumber\\
    \xi=&(r^\mathrm{cw})^{-2} \left(\frac{N}{\operatorname{Var}[w]}-4\right).
\end{align}
Here, $r^\mathrm{cw}$ denotes the correlation coefficient between $C_z$ and $w_z$, where $w_z= \bra{z_U} \hat{W} \ket{z_U}$ with $\hat{W}=\sum_{i=1}^N (1-\hat{\sigma}_i^x)/2$ being the operator of the $x$-basis Hamming weight.
The variance of $w_z$ is denoted by $\operatorname{Var}[w] = \overline{w_z^2} -(\overline{w_z})^2$.
The sum rule, $\overline{\sigma_z^2} + \operatorname{Var}[w] = N/4$, with $\sigma_z =  \sqrt{ \bra{z_U} \hat{W}^2 \ket{z_U} - w_z^2}$, guarantees $\xi\geq 0$ and $A \geq 1$.

Consequently, $\xi$ provides an effective indicator of energy-space locality associated with $d_E^\mathrm{rpa}$, irrespective of whether the underlying cost function exhibits correlations between the Hamming and energy spaces.
A smaller $\xi$ corresponds to stronger localization in energy space.
In the following, we identify how QA and QAOA produce a small $\xi$.

\begin{figure}[t]
    \centering
    \includegraphics[width=0.45\linewidth]{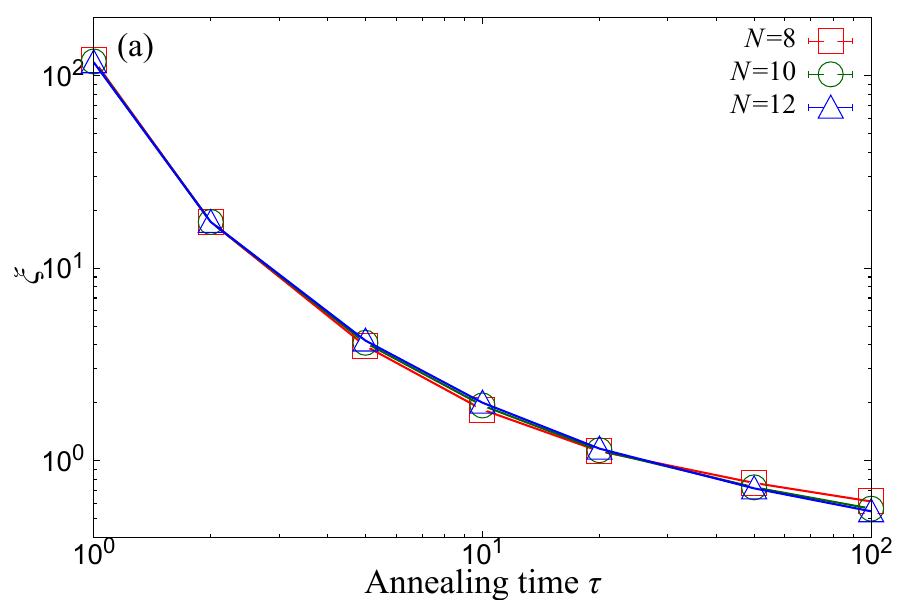}
    \includegraphics[width=0.45\linewidth]{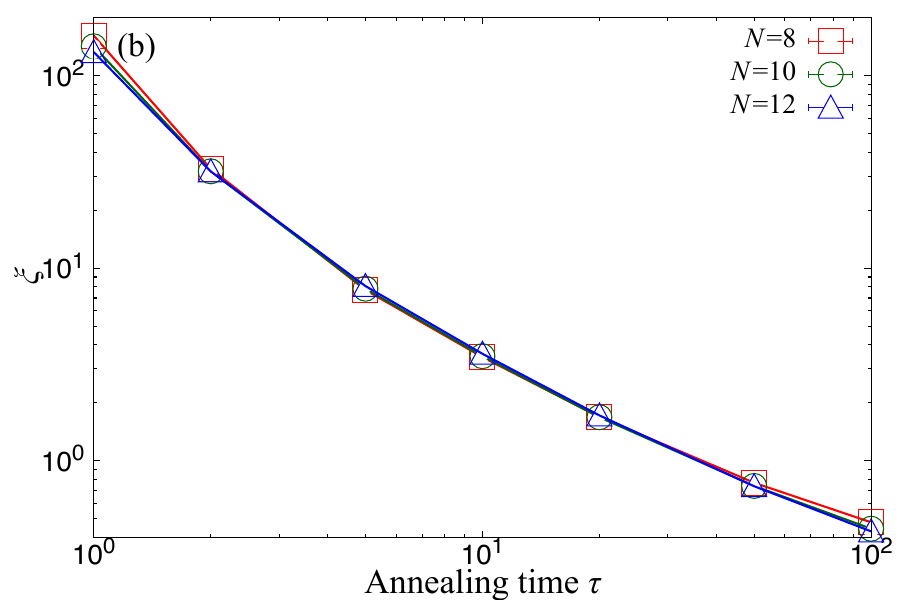}\\
    \includegraphics[width=0.45\linewidth]{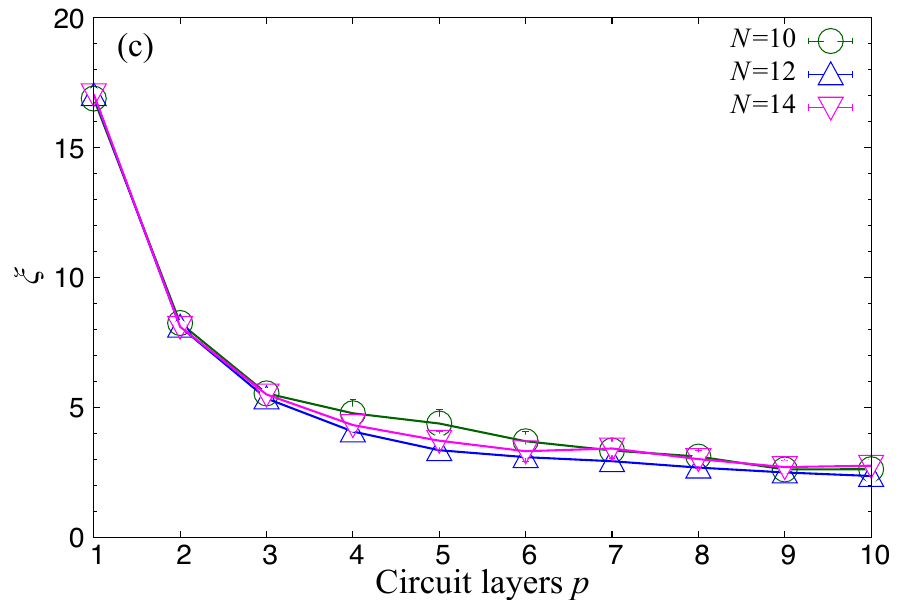}
    \includegraphics[width=0.45\linewidth]{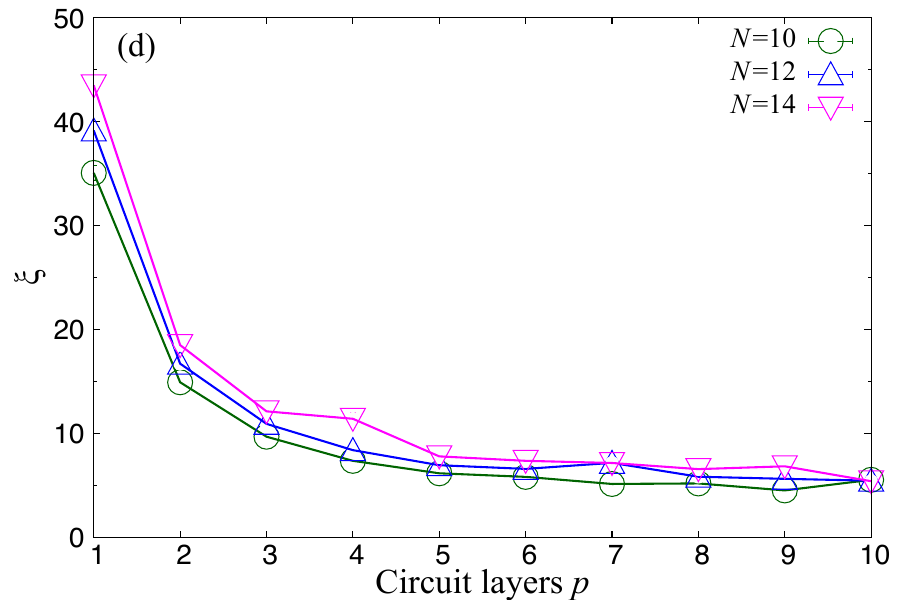}
    \caption{
    Dependences of $\xi$ on the annealing time $\tau$ and the QAOA layers $p$ for (a),(c) the random Ising model and (b),(d) the REM.
    }
    \label{fig:xi}
\end{figure}

For QA, we investigate the $\tau$ dependence.
The upper panels of Fig.~\ref{fig:xi} show that $\xi$ decreases with the annealing time for (a) the random Ising model and (b) the REM.
Since $|r_\mathrm{cw}|$ remains close to unity, the decrease in $\xi$ is mainly attributed to the increase in $\operatorname{Var}[w]$.
In addition, the curves for different system sizes approximately collapse in both models.
These results suggest that the localization in energy space can emerge through finite-time quasiadiabatic dynamics without resolving exponentially small many-body-level spacings.
The weak system-size dependence of $\xi$ implies the scalability of this approach.

For QAOA, the lower panels of Fig.~\ref{fig:xi} show $\xi$ as a function of the number of layers in (c) the random Ising model and (d) the REM.
In both models, $\xi$ decreases with the number of layers.
In the random Ising model, the curves for different system sizes approximately collapse onto a single curve, whereas only weak size dependence is observed in the REM within the accessible system sizes.
As in QA, the decrease in $\xi$ is mainly attributed to the increase in $\operatorname{Var}[w]$, while $|r_\mathrm{cw}|$ remains close to unity.
These results suggest that fixed-depth QAOA can generate energy-resolved ordering even in problems without apparent geometric structures, although its asymptotic scalability, particularly in the REM, remains an open question.

To understand the reduction in $\xi$ with increasing $\tau$ or $p$, we expand
the effective Hamiltonian in the Pauli-string basis as
\begin{equation}
    \hat{C}_\mathrm{eff}=\sum_{\boldsymbol{\mu}} c_{\boldsymbol{\mu}}\hat P_{\boldsymbol{\mu}},
\qquad
\hat P_{\boldsymbol{\mu}}=\bigotimes_{i=1}^{N}\hat\sigma_i^{\mu_i},
\end{equation}
where $\boldsymbol\mu=(\mu_1,\ldots,\mu_N)$, $\mu_i \in \{0,x,y,z\}$, and $c_{\boldsymbol{\mu}} \in \mathbb{R}$.
Here, $\hat{\sigma}_i^0$ is the identity operator.
The expectation value of $\hat{C}_\mathrm{eff}$ in the state $\ket{+}$ can be expressed as
\begin{equation}
    \bra{+} \hat{C}_\mathrm{eff} \ket{+}=\sum_{S \subseteq [N]} c_{\hat{X}_S}, \qquad \hat{X}_S=\prod_{i \in S} \hat{\sigma}_i^x,
\end{equation}
where $c_{\hat{X}_S}$ is the coefficient of $\hat X_S$ in the Pauli-string expansion.
The expectation value typically decreases with increasing $\tau$ in QA and can be reduced by exploiting the additional variational freedom at larger $p$ in QAOA. 
Such a reduction may be accompanied by increasingly negative contributions from single-qubit Pauli-$X$ terms.
To relate these contributions to energy-space localization, we derive (see Appendix~\ref{appendix:derivation} for details)
\begin{equation}
    \operatorname{Var}[w]=\frac{(\sum_{i=1}^N c_{\hat{X}_{\{i\}}})^2}{4 \operatorname{Var}[C]r_\mathrm{cw}^2}.
    \label{eq:var[w]}
\end{equation}
Since $\operatorname{Var}[C]$ is independent of $\tau$ and $p$, and $|r_\mathrm{cw}|$ remains close to unity, an increase in $-\sum_{i=1}^N c_{\hat{X}_{\{i\}}}$ leads to an increase in $\operatorname{Var}[w]$, therefore reducing $\xi$.
This suggests a common mechanism underlying the enhancement of energy-space localization with increasing $\tau$ in QA and $p$ in QAOA and supports the choice of $\hat{V}_i=\hat{\sigma}_i^x$ for promoting energy-space locality.

\section{Application to combinatorial optimization}\label{sec:application}
We next investigate how the locality properties of the quantum-echo Markov process can be exploited for iterative optimization.
We consider two settings.
First, we study quantum-echo local optimization in a ferromagnetic Ising chain and the random Ising model.
The ferromagnetic Ising chain allows us to investigate the algorithm at large system sizes and obtain analytical insights.
The random Ising model has a rugged energy landscape and provides a more generic setting for examining transitions between local minima.
Second, we consider quantum-echo greedy descent in the random Ising model.

Both approaches are hybrid quantum-classical algorithms, in which quantum dynamics generates candidate configurations according to the engineered transition kernel, while classical computation evaluates and updates configurations.
In contrast to the conventional use of variational quantum algorithms~\cite{peruzzo2014variational,mcclean2016theory,cerezo2021variational}, the quantum device is used here to engineer structured transitions between configurations.

\subsection{Quantum-echo local optimization}\label{sec:local}
Here, we consider quantum-echo local optimization, in which a candidate configuration is selected according to $T_{zz'}$ in Eq.~(\ref{eq:transition_probability}),
and then the candidate configuration is accepted when $\Delta C_{zz'}\leq 0$.
These two steps are repeated until a termination condition is satisfied.

\subsubsection{Ferromagnetic Ising chain}
We apply the quantum-echo local optimization to a simple optimization task:  a ferromagnetic Ising chain,
\begin{equation}
    C_z=-\sum_{i=1}^N (-1)^{z_i + z_{i+1}},
\end{equation}
with periodic boundary condition (i.e., $z_{N+1}=z_1$).
This model is mathematically equivalent to the Max-cut model on a one-dimensional ring when $N$ is even.

To investigate large system sizes, we introduce the truncated transition matrix as
\begin{align}
    &T_{zz'}^\mathrm{tr}=\frac{1}{N} \sum_{i=1}^N T^{(i)}_{zz'}+\left(1-\frac{1}{N} \sum_{i=1}^N \sum_{z''} T^{(i)}_{zz''}\right) \delta_{zz'}, \nonumber\\
    &T^{(i)}_{zz'}=\left\{
    \begin{aligned}
        &|\bra{z_U'} \hat{\sigma}_i^x \ket{z_U}|^2 \text{ if } z_j = z'_j \text{ for } j \notin D_{i,\ell}\\
        &0 \text{ otherwise},
    \end{aligned}
    \right.
    \label{eq:approximated_transition_matrix}
\end{align}
with $D_{i,\ell}=\{i-\ell,i-\ell+1, \cdots, i+\ell\}$.
For QAOA, we set $\ell = p$, for which $T_{zz'}^\mathrm{tr}=T_{zz'}$.
For QA, $\ell$ is chosen as the smallest integer satisfying
\begin{equation}
    \frac{2^{-N}}{N}\sum_{z,z'} \sum_{i=1}^N T_{zz'}^{(i)} \geq 0.99,
\end{equation}
when $N=2\ell+2$.
We restrict the annealing time to $\tau \leq 20$, since the required locality range $\ell$ increases with $\tau$, making simulations at longer annealing times computationally intractable.
The value of $\ell$ determined this way is then used for all larger system sizes.
In each iteration of the Markov process, we first choose $i$ uniformly from $\{1,\ldots,N\}$ and then sample a candidate configuration according to $T_{zz'}^{(i)}$.
The missing probability $1-\sum_{z'}T_{zz'}^{(i)}$, resulting from the truncation, is assigned to $z'=z$, consistently with the definition of $T_{zz'}^\mathrm{tr}$.
We use $N=10^6$ for QAOA and $N=10^5$ for QA.
Averages and standard deviations are evaluated on ten simulation runs.

\begin{figure}[t]
    \centering
    \includegraphics[width=0.45\linewidth]{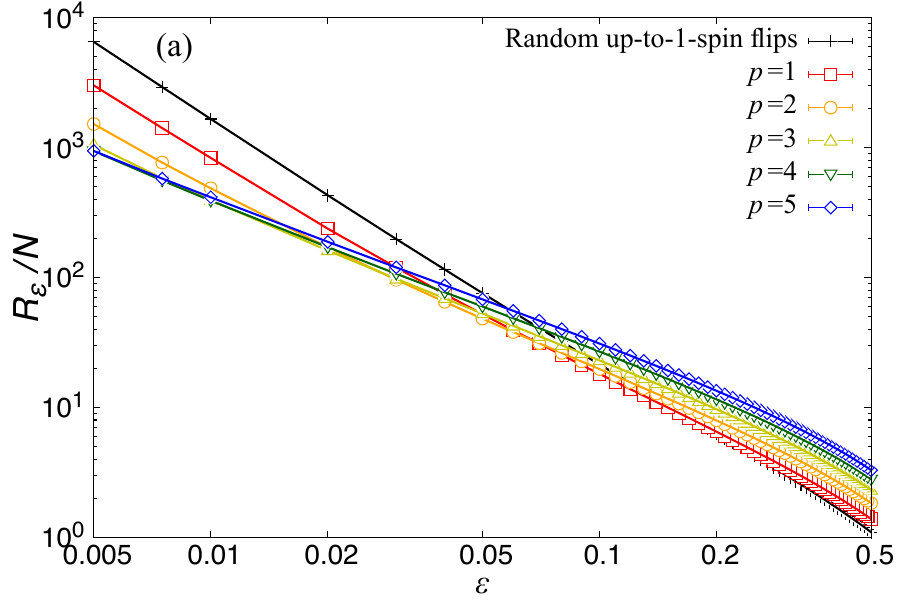}
    \includegraphics[width=0.45\linewidth]{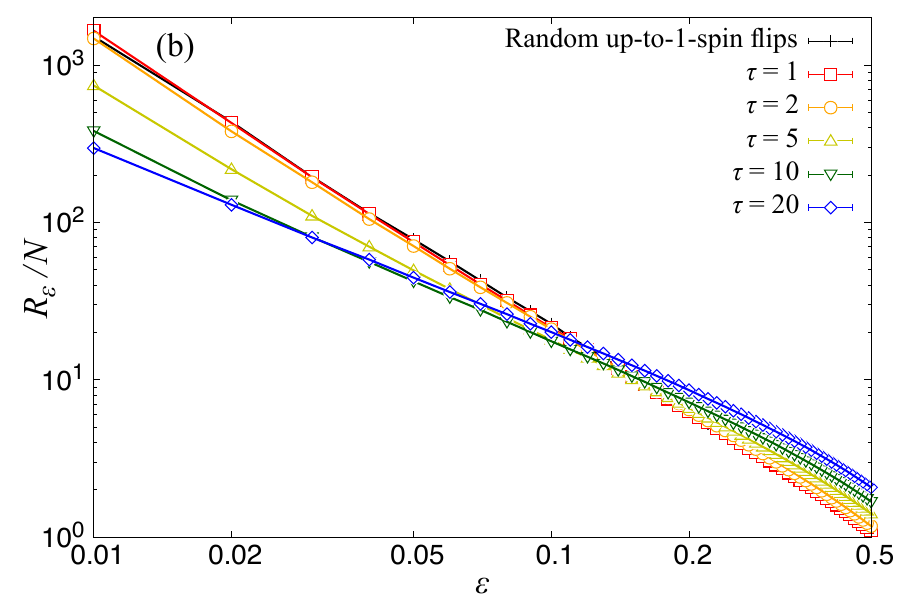}\\
    \includegraphics[width=0.45\linewidth]{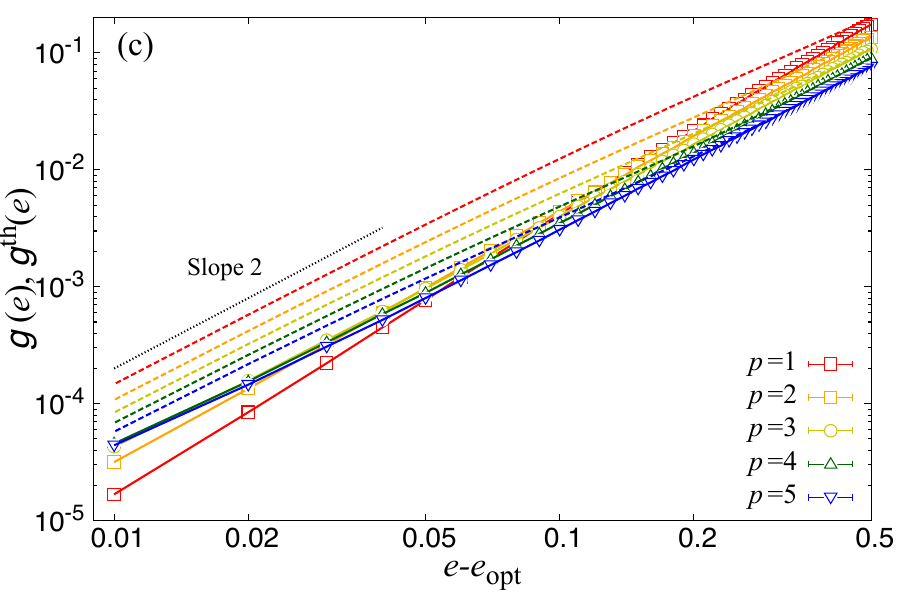}
    \includegraphics[width=0.45\linewidth]{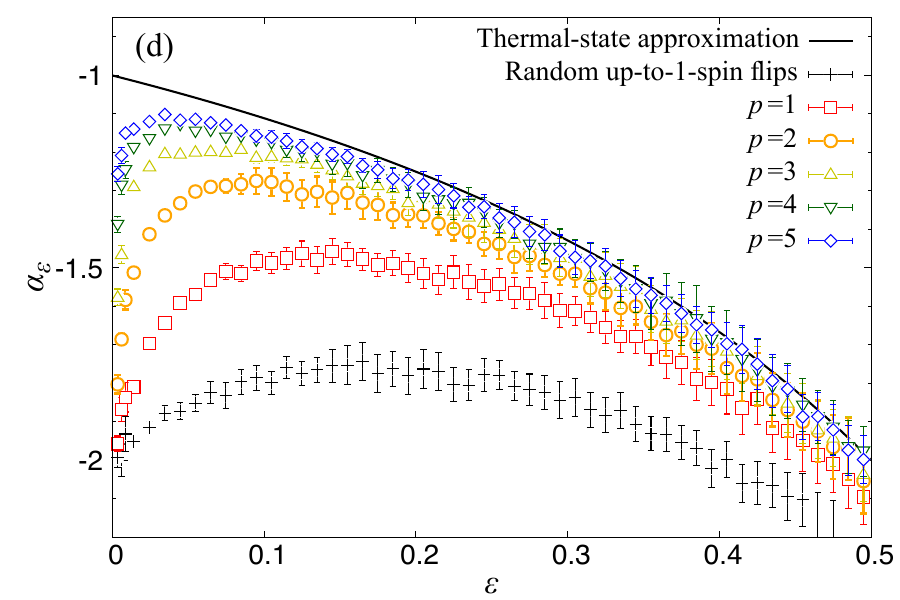}\\
    \includegraphics[width=0.45\linewidth]{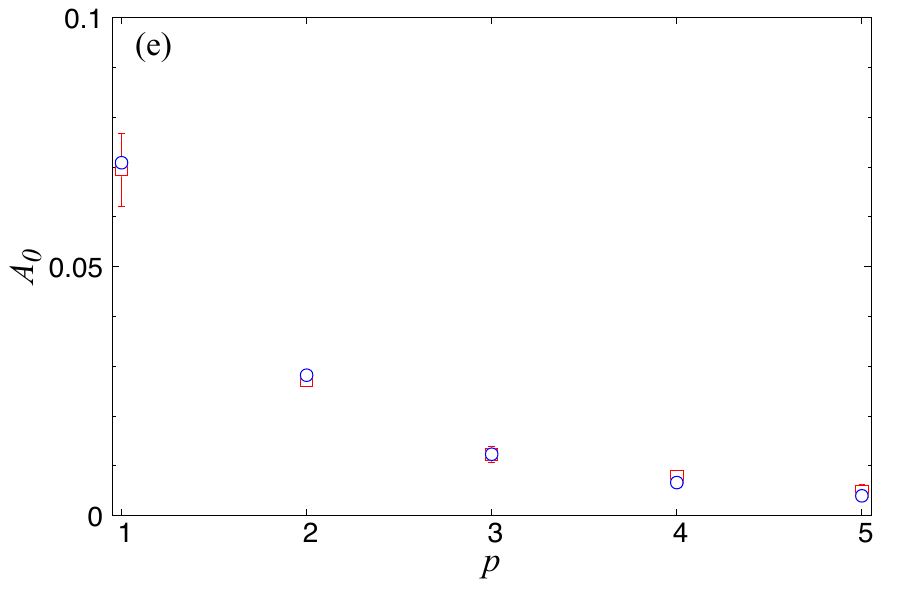}
    \includegraphics[width=0.45\linewidth]{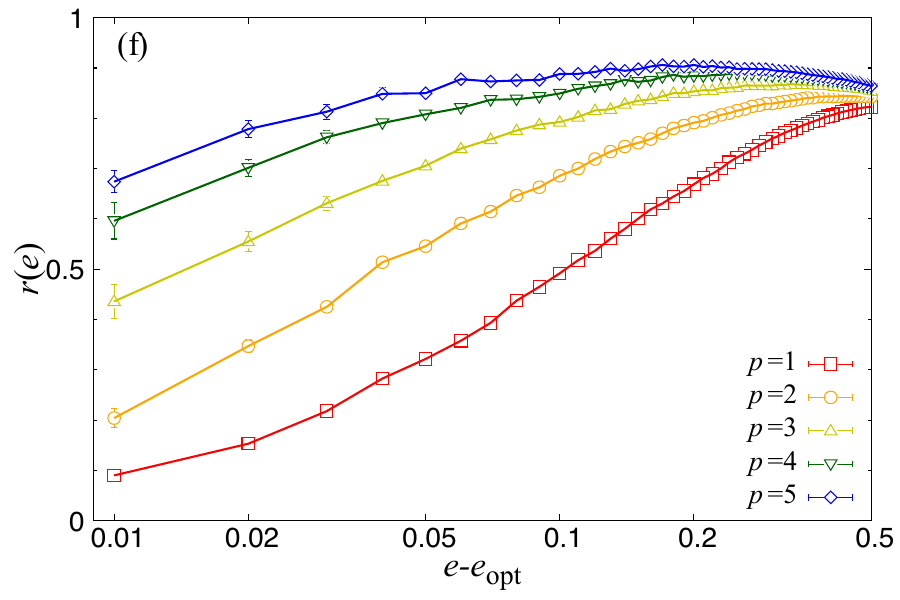}\\
    \includegraphics[width=0.45\linewidth]{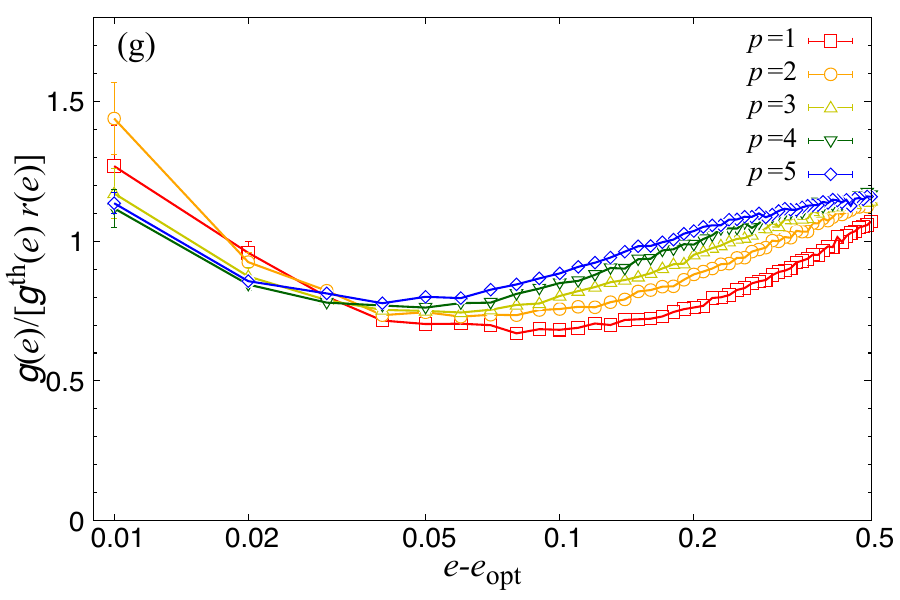}
    \caption{
    Quantum-echo local optimization results for the ferromagnetic Ising chain.
    $R_\epsilon/N$ for different values of (a) the number of layers $p$ and (b) the annealing time $\tau$.
    (c) The dynamical drift $g(e)$ (solid curves) and thermal drift $g^\mathrm{th}(e)$ (dashed curves), (d) the effective exponent $\alpha_\epsilon = d\log(R_\epsilon/N)/d\log\epsilon$, (e) two estimates of the asymptotic amplitude $A_0$ defined by $R_\epsilon \sim N A_0 \epsilon^{-2}$ as $\epsilon \to 0$, (f) local-defect ratio $r(e)$, and (g) the ratio $g(e)/[g^\mathrm{th}(e) r(e)]$.
    }
    \label{fig:Ferromagnetic_chain}
\end{figure}

We evaluate the mean first-passage time $R_\epsilon$, defined as the mean number of iterations required to reach the target optimality gap $\epsilon$.
Figures~\ref{fig:Ferromagnetic_chain}~(a) and~(b) show the $\epsilon$-dependence of $R_\epsilon$ in units of $N$ for different $p$ and $\tau$, respectively.
For comparison, we use random up-to-$1$-spin flips as a baseline.
We find that $R_\epsilon$ scales linearly with $N$ (not shown).
Furthermore, as $\epsilon$ decreases, the optimal layer number and annealing time that minimize $R_\epsilon$ increase.
This result indicates that larger $p$ or $\tau$ becomes advantageous when searching for high-quality configurations.
The quantum-echo transition outperforms the random up-to-$1$-spin flips for small $\epsilon$ when performance is measured by $R_\epsilon$.
The dependence of $R_\epsilon$ is not qualitatively affected by the presence of an exponentially small gap along the annealing path (see Appendix~\ref{appendix:1stQPT}). 

We first rewrite $R_\epsilon$ as
\begin{equation}
    R_\epsilon =-N\int_0^{-(1-\epsilon)} [g(e)]^{-1} de.
    \label{eq:step_to_epsilon}
\end{equation}
Here, $g(e)$ is the drift of total energy per iteration conditioned on the energy density $e$, and is explicitly given as
\begin{equation}
    g(e) \delta e =-\sum_{\substack{z,z'|\Delta C_{zz'} < 0,\\\frac{C_z}{N} \in [e, e+\delta e)}} P_z^\mathrm{dyn}(e) \Delta C_{zz'} T_{zz'},
\end{equation}
where $\delta e$ is a small energy-density width that satisfies $N^{-1} \ll \delta e \ll 1$ and $P_z^\mathrm{dyn}(e)$ denotes the dynamical distribution conditioned on the shell $C_z/N \in [e, e+\delta e)$.
Here, we set $\delta e=0.01$.
Since Hamming-space locality of the transition matrix, together with the cost Hamiltonian locality, ensures an $O(1)$ change in the total energy per iteration, the iteration counts required to traverse an energy-density interval $\delta e$ is approximately $N\delta e/g(e)$, which explains the linear scaling of $R_\epsilon$ with $N$.

We next explain the origin of the crossover of the optimal number of layers and the annealing time.
The following discussion focuses on QAOA, as QA exhibits qualitatively the same behavior with respect to $\tau$ as QAOA does with respect to $p$.
Figure~\ref{fig:Ferromagnetic_chain}~(c) shows $g(e)$ for different $p$.
$g(e)$ decreases as the energy density approaches $e_\mathrm{opt}=-1$.
For comparison, we introduce the energy drift in the thermal state:
\begin{equation}
    g^{\mathrm{th}}(e)=- \sum_{z,z'|\Delta C_{zz'} < 0} P_z^\mathrm{th}(\beta) \Delta C_{zz'} T_{zz'},
\end{equation}
where $P_z^\mathrm{th}(\beta) = e^{-\beta C_z}/\sum_{z'}e^{-\beta C_{z'}}$ with $\beta$ determined by $\sum_z C_z P_z^\mathrm{th}(\beta)=Ne$.
Although the thermal drift $g^\mathrm{th}(e)$ reproduces the dynamical drift $g(e)$ well at high energies, the deviation increases as the energy approaches the optimum.
The $p$-dependence in the low-energy regime is qualitatively different between $g(e)$ and $g^\mathrm{th}(e)$: $g(e)$ increases with $p$, while $g^\mathrm{th}(e)$ decreases with $p$.
The thermal drift scales near the optimum independently of $p$ as
\begin{equation}
    g^\mathrm{th}(e) \sim (e-e_\mathrm{opt})^2.
    \label{eq:thermal_drift}
\end{equation}
This is because energy improvement requires eliminating a pair of nearby domain walls.
Since the density of domain walls in thermal states is proportional to the energy difference from the optimum, the probability of finding such a pair scales quadratically with the energy-density difference from the optimum.

Figure~\ref{fig:Ferromagnetic_chain}~(d) shows the effective exponent $\alpha_\epsilon=d \log(R_\epsilon /N)/d\log \epsilon$ for different $p$.
$\alpha_\epsilon$ generally increases with $p$ and exhibits a nonmonotonic dependence on $\epsilon$.
As $\epsilon$ decreases, $\alpha_\epsilon$ initially increases and then decreases near the optimum. 
The behavior at relatively large $\epsilon$ can be  qualitatively explained by the thermal-state approximation.
Assuming the quadratic scaling of $g^\mathrm{th}(e)$ in Eq.~(\ref{eq:thermal_drift}) remains valid over the entire energy range, Eq.~(\ref{eq:step_to_epsilon}) gives an effective exponent as
\begin{equation}
    \alpha_\epsilon^\mathrm{th}=-\frac{1}{1-\epsilon}.
\end{equation}
The solid line in Figure~\ref{fig:Ferromagnetic_chain}~(d) represents this prediction and qualitatively reproduces the behavior at relatively large $\epsilon$.
As $\epsilon$ approaches zero, however, $\alpha_\epsilon$ deviates from this prediction and decreases.
For small $p$, $\alpha_\epsilon$ goes toward $-2$, which is expected from diffusion-annihilation dynamics of domain walls~\cite{spouge1988exact,amar1990diffusion}.
Although the accessible range of $\epsilon$ is insufficient to establish the asymptotic exponent directly for larger $p$, the decrease in $\alpha_\epsilon$ near the optimum is consistent with this prediction.
These observations suggest a crossover from a thermal-state-like regime to a diffusion-annihilation regime.
Importantly, increasing $p$ extends the thermal-state-like regime toward smaller $\epsilon$, yielding an effective exponent to $-1$ at the crossover.
This provides a nearly quadratic speedup in the $\epsilon$ dependence of the mean first-passage time compared with random up-to-single spin flips.

Figure~\ref{fig:Ferromagnetic_chain}~(e) compares two estimates of the asymptotic amplitude $A_0$, defined by $R_\epsilon/N \sim A_0 \epsilon^{-2}$ (see details in Appendix~\ref{appendix:extrapolation}).
The red squares represent estimates obtained by extrapolating $\epsilon^2 R_\epsilon/N$ to $\epsilon =0$.
The blue circles represent the theoretical prediction $1/(2\pi D)$ for diffusion-annihilation dynamics~\cite{spouge1988exact}, where $D$ is the diffusion coefficient of an isolated domain wall.
The two estimates agree within the uncertainties for all values of $p$ considered, supporting the interpretation that the asymptotic low-energy dynamics is governed by diffusion-controlled annihilation of domain walls.
The decrease in $A_0$ with increasing $p$ indicates that quantum-echo dynamics accelerates the approach to the optimum by reducing the prefactor of $R_\epsilon$, while preserving the $\epsilon^{-2}$ scaling.

To understand the mechanism behind the $p$ dependence, we next compare local structures between the dynamical state and the thermal state.
The local search can only modify local structures and thus generally drives the dynamical state away from the thermal state.
To quantify this effect, we define the local-defect ratio $r(e)$ as the local-defect density in the dynamical state divided by that in the thermal state at the same energy density.
Here, local defects are defined as the sequence of spins $(0,1,0)$, $(1,0,1)$, $(0,1,1,0)$, and $(1,0,0,1)$, respectively.
The total local-defect density in the thermal state is given by $(e-e_\mathrm{opt})^2/2$.
Figure~\ref{fig:Ferromagnetic_chain}~(f) shows that $r(e)$ remains below unity in the energy range considered.
It approaches unity as the number of layers increases, whereas it decreases as the energy approaches the optimum.
Consistently with the behavior of $\alpha_\epsilon$, this result indicates that the thermal-state approximation is more accurate as the number of layers increases.

Figure~\ref{fig:Ferromagnetic_chain}~(g) plots $g(e)/[g^\mathrm{th}(e)r(e)]$ for different $p$.
Despite systematic deviations from unity, the ratio remains approximately within the range $0.6-1.5$ in the energy range considered.
This indicates that the dynamical drift is approximately described by
\begin{equation}
    g(e) \approx g^{\mathrm{th}}(e) r(e).
\end{equation}
The remaining deviations indicate dynamical correlations that are not captured solely by the local-defect ratio.
For small $p$, $r(e)$ decreases substantially as the energy approaches the optimum, thus suppressing the dynamical drift relative to the thermal drift.
In contrast, for large $p$, $r(e)$ remains closer to unity in the low-energy regime, resulting in weaker suppression of the dynamical drift and allowing deeper circuits to reach smaller optimality gaps in fewer iterations.

Then, we compare quantum-echo local optimization with random up-to-$k$-spin flips.
For fixed $k$ and large $N$, transitions involving exactly $k$ spin flips dominate the random up-to-$k$-spin flip process.
The thermal drift near the optimum therefore scales as
\begin{equation}
    g^\mathrm{th}(e)\sim (e-e_\mathrm{opt})^{k+1}
\end{equation}
because, in the lowest order, an energy improvement requires eliminating a domain-wall pair while shifting other $k-1$ domain walls.
The scaling exponent increases with $k$, resulting in the rapid decay of the thermal drift for large $k$.
Unlike random $k$-spin flips, increasing $p$ in the quantum-echo Markov process preserves the quadratic scaling of the thermal drift.
This is a key advantage of quantum-echo local optimization over random up-to-$k$ spin flips.

Finally, for comparison, when QA and QAOA are used as standalone optimizers, achieving an optimality gap $\epsilon$ requires an annealing time of $O(\epsilon^{-2})$ for QA~\cite{dziarmaga2005dynamics} and $O(\epsilon^{-1})$ layers for QAOA~\cite{mbeng2019quantum}.
In contrast, in an iterative local optimization process, a fixed finite-time or finite-depth transition rule can be repeatedly applied to progressively reduce the optimality gap.

\subsubsection{Random Ising model}\label{subsec:RIM}
\begin{figure}[t]
    \centering
    \includegraphics[width=0.45\linewidth]{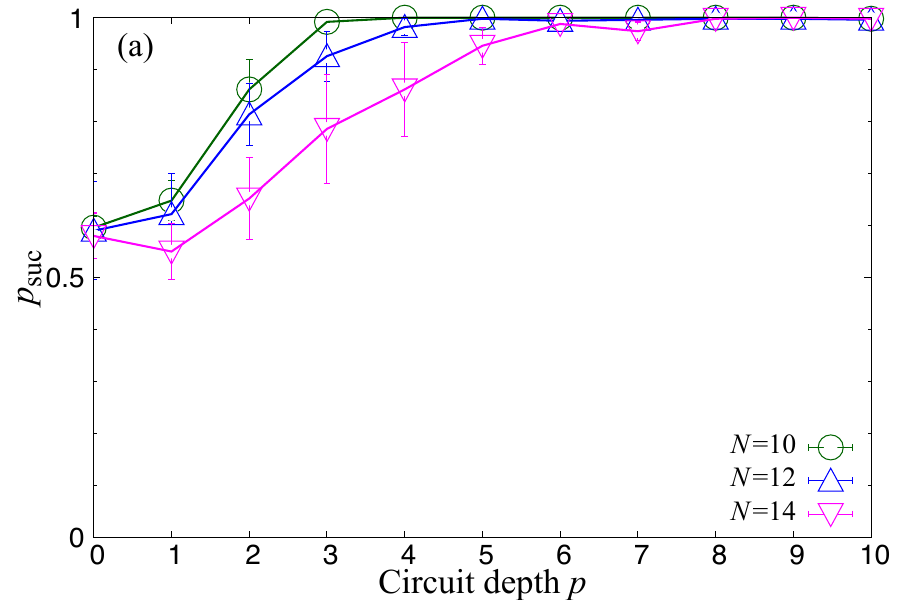}
    \includegraphics[width=0.45\linewidth]{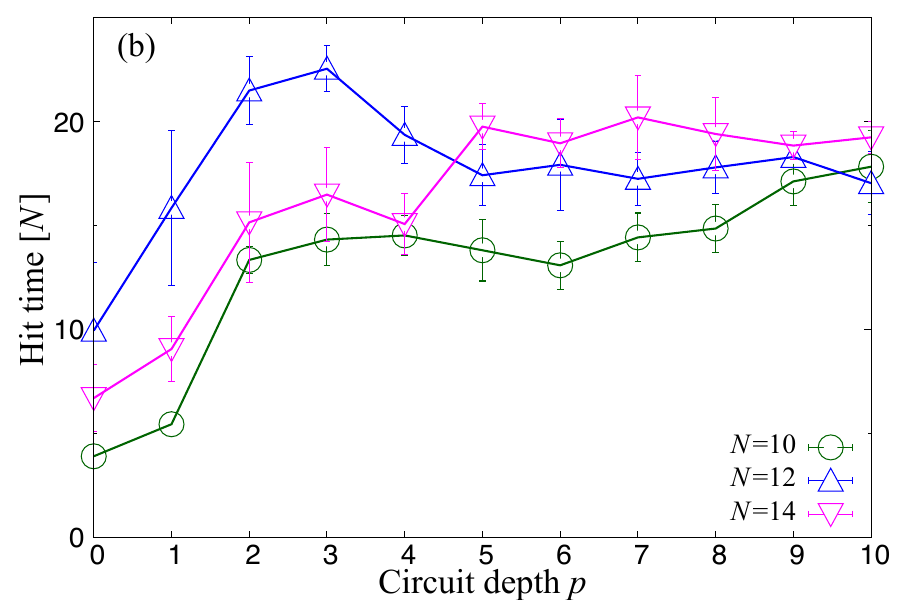}\\
    \includegraphics[width=0.45\linewidth]{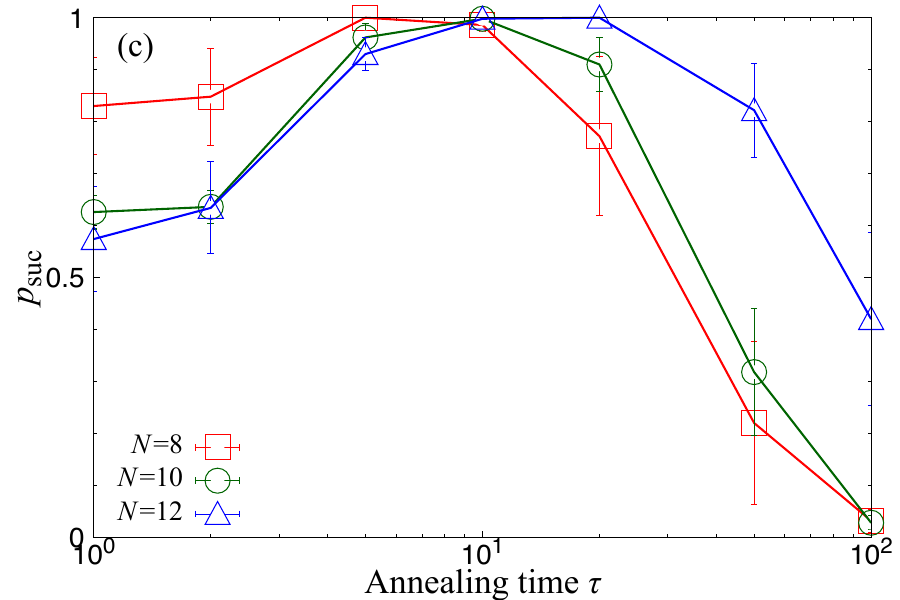}
    \caption{
    Quantum-echo local optimization results for the random Ising model.
    (a) Success probability $p_\mathrm{suc}$ and (b) hit time in units of $N$ as functions of the QAOA layer number $p$.
    (c) $p_\mathrm{suc}$ as a function of the QA annealing time $\tau$.
    }
    \label{fig:local}
\end{figure}
We next apply the quantum-echo local optimization to the random Ising model.
For performance evaluation, we examine the success probability $p_\mathrm{suc}$ and the hit time.
$p_\mathrm{suc}$ is defined as the probability of reaching one of the optimal configurations within $100N$ iterations on $100$ trials.
The hit time is defined as the average iteration counts required to reach an optimum solution, conditioned on successful trials.

Figures~\ref{fig:local}~(a) and~(b) show the results for quantum-echo local optimization based on QAOA.
$p_\mathrm{suc}$ increases with the number of circuit layers $p$, while the increase becomes more gradual as the system size increases.
The initial increase in hit time is mainly due to the inclusion of trials that fail to reach an optimum at a smaller $p$.
These newly successful trials typically require a large iteration counts before reaching an optimum.
Importantly, once $p_\mathrm{suc}$ approaches unity, the hit time becomes nearly independent of $p$.

For quantum-echo local optimization based on QA, Fig.~\ref{fig:local}~(c) shows a nonmonotonic behavior of $p_\mathrm{suc}$.
It initially increases with the annealing time $\tau$, but decreases toward zero for large $\tau$.
In the adiabatic limit, the transition matrix becomes block diagonal, so that quantum-echo local optimization cannot reach an optimal solution unless the initial state belongs to a block containing one. 

These results demonstrate that stronger energy-space localization does not necessarily improve the performance and can suppress the transitions required to reach an optimum in the extreme case.

\subsection{Quantum-echo greedy descent}\label{sec:basin-hopping}
The results in Sec.~\ref{subsec:RIM} show that increasing $p$ or $\tau$ in the quantum-echo transition does not necessarily improve quantum-echo local optimization.
We therefore consider an enhanced strategy that combines the quantum-echo transitions with greedy descent based on single-spin flips in each iteration.
The greedy descent repeatedly chooses one of the strictly downhill single-spin flips uniformly at random until a local minimum is reached.
Here, a local minimum is defined as a state for which no single-spin flip decreases the cost function, and $\mathcal{D}^\mathrm{L}$ denotes the set of local minima.
The transition probability between local minima $z, z' \in \mathcal{D}^\mathrm{L}$ is then defined as
\begin{equation}
    T^\mathrm{L}_{zz'}=\sum_{z''} T_{zz''} T_{z''z'}^\mathrm{greedy},
    \label{eq:transition_local}
\end{equation}
where $T_{zz'}^\mathrm{greedy}$ is the probability that the greedy descent starting from $z$ terminates at $z' \in D^\mathrm{L}$.

Algorithm~\ref{Algo:basin} summarizes quantum-echo greedy descent.
Starting from an initial local minimum, the algorithm iteratively generates a candidate local minimum according to the transition probability $T^\mathrm{L}_{zz'}$.
The candidate is accepted if its energy does not exceed that of the current local minimum.
This procedure is repeated until a stopping criterion is reached.
Here, we set the maximum iteration counts to $100N$.

\begin{algorithm}[t]
\caption{Quantum-echo greedy descent}
\label{Algo:basin}

\KwIn{$z_\mathrm{ini}$}
\KwOut{$z_\mathrm{best}$}

Locally optimize $z_\mathrm{ini}$ to a local minimum $z$\;
$z_\mathrm{best} \gets z$

\While{stopping criterion is not satisfied}{
    Generate $z'$ according to $T_{zz'}^\mathrm{L}$\;
    \If{$C_{z'} \le C_z$}{
        $z \gets z'$\;
    }
}
$z_\mathrm{best} \gets z$\;
\end{algorithm}

\begin{figure}[t]
    \centering
    \includegraphics[width=0.45\linewidth]{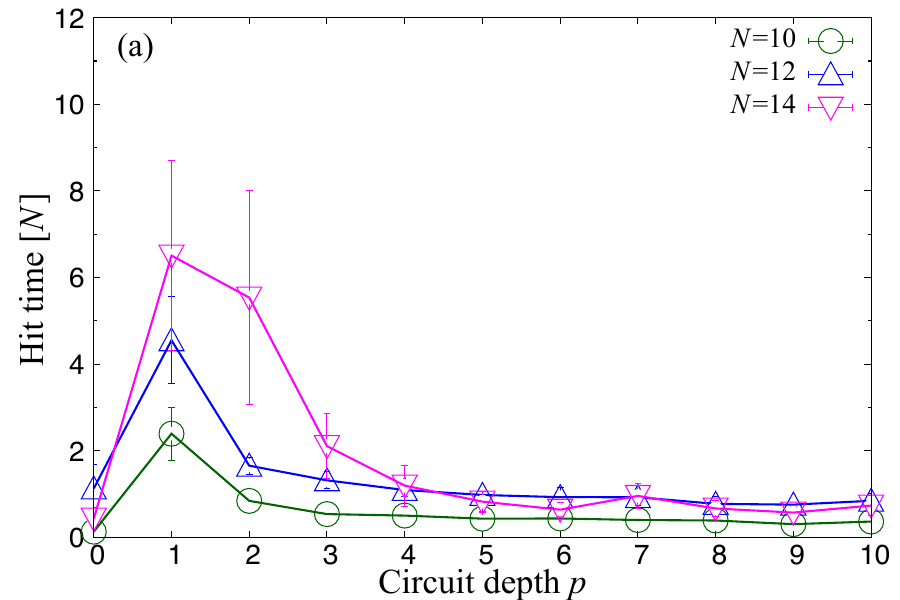}
    \includegraphics[width=0.45\linewidth]{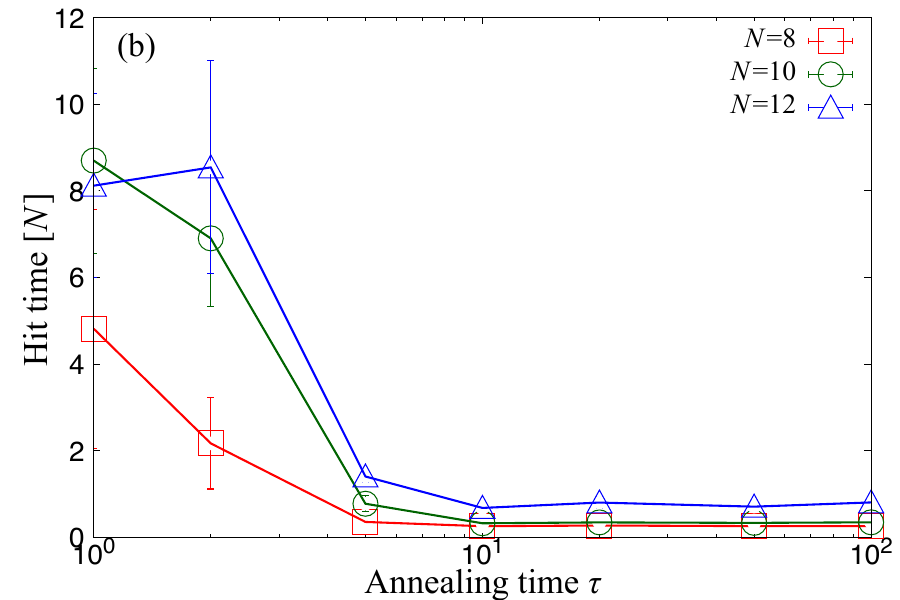}\\
    \includegraphics[width=0.45\linewidth]{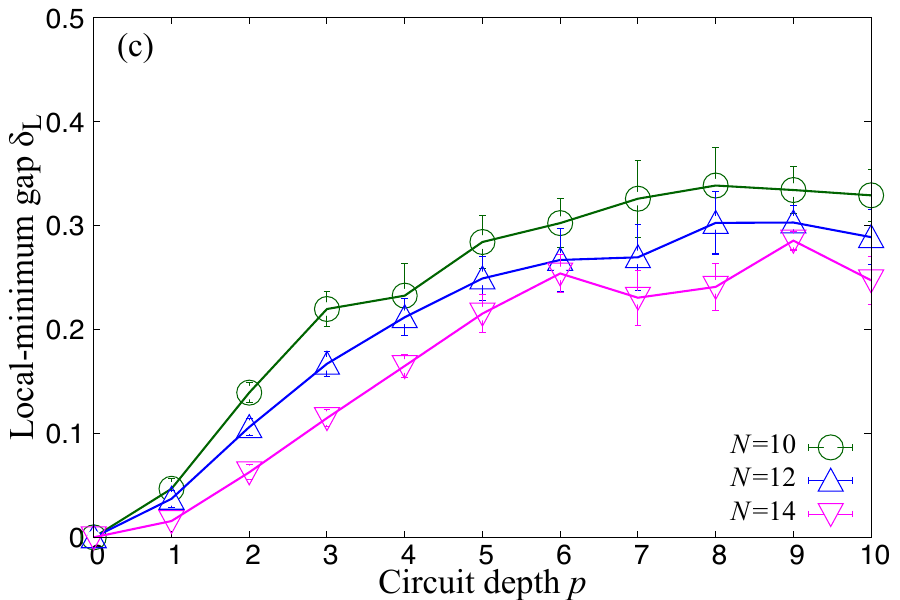}
    \includegraphics[width=0.45\linewidth]{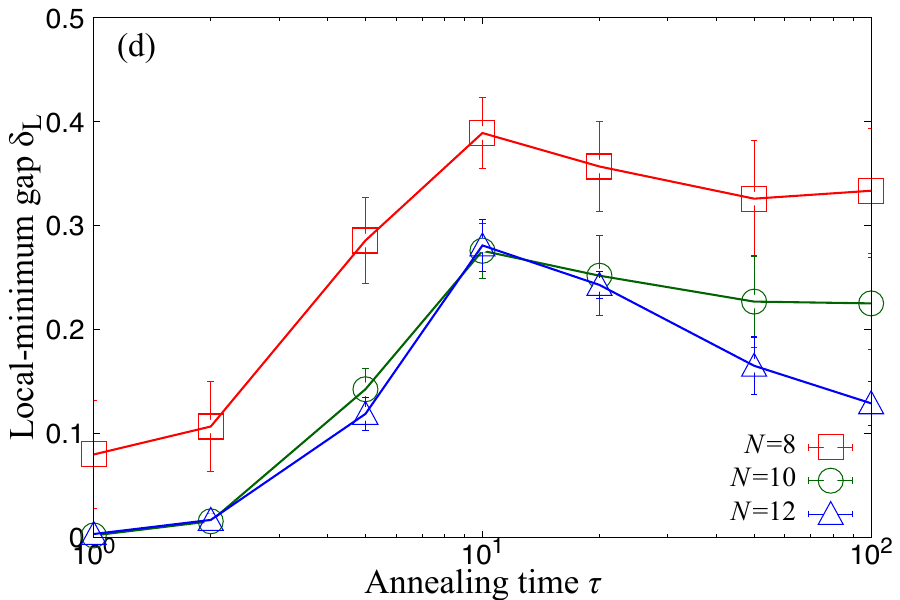}
    \caption{
    Results of quantum-echo greedy descent for the random Ising model.
    Hit time in units of $N$ as functions of (a) the number of QAOA layers $p$ and (b) the annealing time $\tau$.
    Local-minimum gap as functions of (c) $p$ and (d) $\tau$.
    }
    \label{fig:basin_hopping}
\end{figure}

We find that $p_\mathrm{suc}$ rapidly approaches unity with smaller $p$ or $\tau$ than in quantum-echo local optimization.
More importantly, for QA, the success probability remains unity even at large $\tau$, where quantum-echo local optimization becomes ineffective.
Figures~\ref{fig:basin_hopping}~(a) and~(b) show that the hit time decreases with increasing $p$ or $\tau$.

To characterize transitions among local minima, we define the local-minimum gap as 
\begin{equation}
 \delta_\mathrm{L}=1-|\lambda_2|,
\end{equation}
where $\{\lambda_i \}$ are the eigenvalues of $T^\mathrm{L}_{zz'}$, ordered as $1=\lambda_1\geq |\lambda_2| \geq \ldots$.
The local-minimum gap characterizes the asymptotic relaxation of the Markov process among local minima toward its stationary distribution, with a larger $\delta_\mathrm{L}$ corresponding to faster asymptotic relaxation.
Figures~\ref{fig:basin_hopping}~(c) and~(d) show that $\delta_\mathrm{L}$ exhibits an overall increasing trend with $p$, while its dependence on $\tau$ is nonmonotonic: it increases at short $\tau$, reaches a maximum at an intermediate $\tau$, and decreases at longer $\tau$, as is particularly evident for $N=12$.
The reduction in the hit time from short to intermediate $\tau$ is therefore accompanied by enhanced mixing among the local minima.
At longer annealing times, however, the decrease in $\delta_\mathrm{L}$ is not accompanied by a corresponding increase in the hit time, suggesting that the asymptotic relaxation characterized by $\delta_\mathrm{L}$ does not necessarily control the access to the optimum.

The contrasting behaviors of quantum-echo local optimization and quantum-echo greedy descent highlight the complementary roles of quantum and classical dynamics.
The quantum-echo transition enables nonlocal exploration of the configuration space, while the subsequent greedy descent drives the configuration toward a nearby local minimum.

\section{Conclusion}\label{sec:conclusion}
We propose a quantum-echo Markov process based on QA and QAOA.
For both the random Ising model and the REM, increasing the annealing time in QA or the number of layers in QAOA yields transitions that are nonlocal in Hamming space but local in energy space.
Hamming-space nonlocality originates from operator-size growth, whereas energy-space locality is associated with the correlation between $C_z$ and $w_z$.
The emergence of this correlation may be related to the decrease in $\bra{+} \hat{C}_\mathrm{eff} \ket{+}$ with increasing $\tau$ in QA or $p$ in QAOA.

For optimization, we introduce quantum-echo local optimization.
For the ferromagnetic Ising chain, the mean first-passage time approaches the scaling $R_\epsilon \sim N \epsilon^{-1}$ as $\tau$ in QA or $p$ in QAOA increases, representing a nearly quadratic improvement in its dependence on $\epsilon^{-1}$ over random single-spin flips, for which $R_\epsilon \sim N \epsilon^{-2}$.
For the random Ising model, increasing $\tau$ or $p$ initially improves optimization performance through the interplay between Hamming-space nonlocality and energy-space locality.
However, further increases lead to performance degradation in QA and saturation in QAOA, suggesting that excessive energy-space locality does not necessarily improve optimization performance.
Quantum-echo greedy descent substantially improves optimization performance, highlighting the complementary roles of quantum dynamics in exploration and greedy descent in exploitation.

Our numerical results for the random Ising model suggest that finite-time quantum dynamics can generate transition kernels with controlled locality properties that persist with increasing system size.
In particular, $\xi$ remains nearly independent of system size.
It would be interesting to characterize such transition kernels directly in the thermodynamic limit and determine whether their observed Hamming- and energy-space structures persist.
For QAOA, an additional question is whether the parameters that generate such kernels can be efficiently learned at large system sizes, where trainability issues such as a barren plateau may arise.
From this perspective, investigating nonvariational parameter schedules derived from, for example, QA protocols would also be useful.

For optimization, the optimal annealing time and number of layers in terms of $R_\epsilon$ for the ferromagnetic Ising chain remain nearly independent of the system size, indicating that finite-time quantum dynamics is sufficient to generate useful transition kernels for this problem.
Whether this favorable behavior persists for harder problems remains an open question.
It is therefore important to investigate the computational scaling of quantum-echo-assisted search when the quantum evolution time is allowed to grow polynomially with the problem size, rather than restricting attention to finite-time quantum dynamics.

Furthermore, quantum-echo local optimization considered here rejects all moves that increase the cost.
More sophisticated acceptance rules, such as those allowing occasional uphill moves, may further improve the performance.
Since our numerical analysis of the random Ising model is restricted to relatively small system sizes, systematic investigation of such acceptance rules and their scaling behavior is left for future work.

Finally, applying the quantum-echo Markov process to sampling thermal equilibrium distributions and comparing its performance with classical MCMC and existing quantum-enhanced MCMC are important directions for future work.

\begin{acknowledgments}
This work was supported by JSPS KAKENHI (Grant Number 26K14859).
The numerical calculations were partly supported by the supercomputer center of ISSP of Tokyo University.
\end{acknowledgments}

\appendix
\twocolumngrid

\section{Special case in random Ising model}\label{appendix:SC_RIM}
The case where the transition matrix of the quantum-echo Markov process becomes $T_{zz'}=\delta_{zz'}$ should be omitted, even if the variational optimization in QAOA minimizes $\bra{+} \hat{C}_\mathrm{eff} \ket{+}$.

We find that this case can appear in the random Ising model, defined in the main text.
When the variational parameters are chosen as $(\gamma_1, \gamma_2, \beta_1, \beta_2)=(\pi/8, -\pi/8, \pm \pi/2, \pi/4)$ or $(-\pi/8, \pi/8, \pm \pi/2, -\pi/4)$ at $p=2$,
\begin{equation}
    \hat{U}^\dagger \hat{\sigma}_i^z \hat{U} =h_i \hat{\sigma}_i^x.
\end{equation}
As a result, $T_{zz'}=\delta_{zz'}$ and the energy expectation value yields
\begin{equation}
    \bra{+} \hat{C}_\mathrm{eff} \ket{+}=-\sum_{(i,j)\in E} J_{ij}h_i h_j -N.
\end{equation}
This choice of variational parameters can appear when the interaction terms have a small contribution compared to the magnetic-field terms.

This example represents the extreme limit of energy-space localization.
Although $\xi$ vanishes, the search process freezes because the transitions between distinct computational-basis states are completely suppressed.
Therefore, $\xi$ alone should not be regarded as a measure of optimization performance.
Instead, efficient optimization requires both sufficient Hamming-space delocalization and an appropriate degree of energy-space locality.

\section{Derivations of $d_E^\mathrm{rpa}$ and $d_E^\mathrm{cw}$}~\label{appendix:derivation}
We derive $d_E^\mathrm{rpa}$ and $d_E^\mathrm{cw}$ through a sequence of approximations.
We first introduce the random-phase approximation.
The transition matrix can be written as
\begin{equation}
    T_{zz'}=\frac{1}{N} \sum_{i=1}^N \sum_{x,x'} (-1)^{x_i+x'_i} G_{zz'}^{x} (G_{zz'}^{x'})^*,
\end{equation}
where
\begin{equation}
    G_{zz'}^x =\langle{z'_U}\ket{x} \langle{x}\ket{z_U}.
\end{equation}
In the random-phase approximation, we neglect the off-diagonal contributions in $x$ and $x'$ by replacing $G_{zz'}^{x} (G_{zz'}^{x'})^*$ by $|G_{zz'}^{x}|^2 \delta_{xx'}$.
The resulting transition matrix is
\begin{equation}
    T_{zz'}^\mathrm{rpa}=\sum_x |G_{zz'}^{x}|^2 = \sum_{x} p_z^x p_{z'}^x.
\end{equation}

We next introduce the shell-uniformity approximation.
We assume that $p_z^x$ is uniform within each Hamming-weight shell specified by $n=\sum_{i=1}^N x_i$.
That is, we approximate $p_z^x$ by $p_z^n/\tbinom{N}{n}$, where $p_z^n=\sum_{x|\sum_{i=1}^N x_i=n}p_z^x$, leading to the transition matrix:
\begin{equation}
    T_{zz'}^\mathrm{shell} =\sum_{n=0}^N \frac{p_z^n p_{z'}^n}{\binom{N}{n}}.
    \label{eq:shell}
\end{equation}

We then approximate the distribution over the Hamming-weight shells by a Gaussian,
\begin{equation}
    p_z^n \approx f_\mathrm{G}(n;w_z,\overline{\sigma_z^2}),
\end{equation}
where $f_\mathrm{G}(x;\mu,\sigma^2)=\frac{1}{\sqrt{2\pi\sigma^2}} \exp (-(x-\mu)^2/2\sigma^2)$.
Here, the mean and variance are determined from the shell distribution as $w_z=\sum_{n=0}^N n p_n^z$ and $\sigma_z^2=(\sum_{n=0}^N n^2 p_n^z)-w_z^2$.
We further assume that the variance is approximately independent of $z$, $\sigma_z^2 \approx \overline{\sigma_z^2}$.
For large $N$, using Stirling's approximation for the binomial coefficient, Eq.~(\ref{eq:shell}) becomes
\begin{equation}
    T_{zz'}^\mathrm{shell} \approx \frac{N^{3/2}}{(2\pi)^{1/2} \overline{\sigma_z^2}} \int_0^1 dx \sqrt{x(1-x)} \exp [-N I_{zz'}(x)],
\end{equation}
where
\begin{align}
    I_{zz'}(x) =&\frac{(x-w_z/N)^2+(x-w_{z'}/N)^2}{2\overline{\sigma_z^2}/N} \nonumber\\
    &-\left[ x \log x +(1-x) \log(1-x)\right].
\end{align}
Since $\overline{w_z}=N/2$ and $\operatorname{Var}[w] \sim N$, $I_{zz'}(x)$ is minimized near $x=1/2$ for typical pairs of $z$ and $z'$.
Expanding the integrand around $x=1/2$ then gives
\begin{align}
    T_{zz'}^\mathrm{shell} \approx T_{zz'}^\mathrm{g1}= &\frac{1}{2^N} \sqrt{\frac{\pi}{2}} \frac{N}{\sqrt{N-2\overline{\sigma_z^2}}} f_\mathrm{G}(w_{zz'};0,2\overline{\sigma_z^2}) \nonumber\\
    &\times \exp\left[\frac{(w_z+w_{z'}-N)^2}{2(N-2 \overline{\sigma_z^2})}\right],
\end{align}
where $w_{zz'}=w_{z'}-w_z$.
We further neglect the fluctuations of $w_z+w_{z'}$ around its typical value $N$, which yields
\begin{equation}
    T_{zz'}^\mathrm{g2} = \frac{1}{2^N} \sqrt{\frac{\pi}{2}} \frac{N}{\sqrt{N-2\overline{\sigma_z^2}}} f_\mathrm{G}(w_{zz'};0,2\overline{\sigma_z^2}).
\end{equation}

Finally, we relate the Hamming-weight difference $w_{zz'}$ to the cost difference $\Delta C_{zz'}$.
A linear least-squares regression gives $w_z \approx a C_z+b$, and hence
\begin{equation}
    w_{zz'} \approx a \Delta C_{zz'}.
\end{equation}
Using
\begin{equation}
    a=\frac{\operatorname{Cov}[C,w]}{\operatorname{Var}[C]} \text{ and } r^\mathrm{cw} = \frac{\operatorname{Cov}[C,w]}{\sqrt{\operatorname{Var}[C]\operatorname{Var}[w]}},
\end{equation}
where $\operatorname{Cov}[C,w]$ is the covariance of $C_z$ and $w_z$, we obtain
\begin{equation}
    T_{zz'}^\mathrm{cw} = \frac{A}{2^N} \exp \left( - \frac{\Delta C_{zz'}^2}{\xi \operatorname{Var}[C]} \right).
\end{equation}
Substituting the approximate transition matrix into the definition of $d_E$ gives $d_E^\alpha$ for $\alpha \in \{\mathrm{rpa}, \mathrm{shell}, \mathrm{g1}, \mathrm{g2}, \mathrm{cw} \}$.

\begin{figure}[t]
    \centering
    \includegraphics[width=0.45\linewidth]{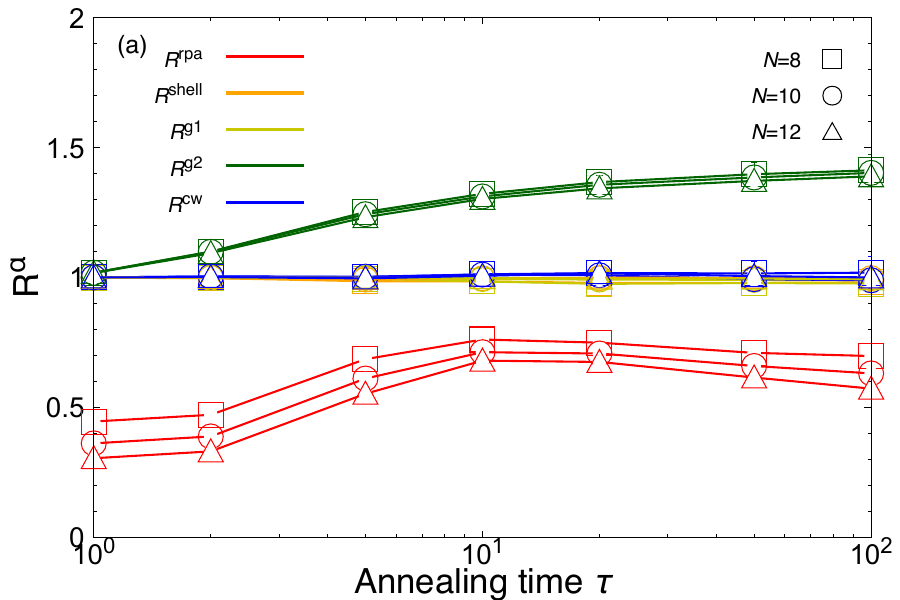}
    \includegraphics[width=0.45\linewidth]{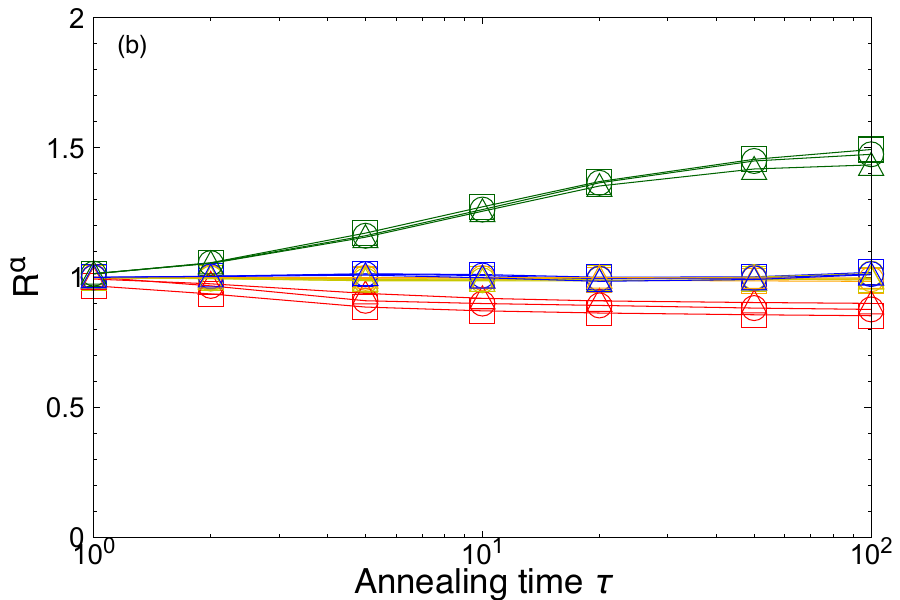}\\
    \includegraphics[width=0.45\linewidth]{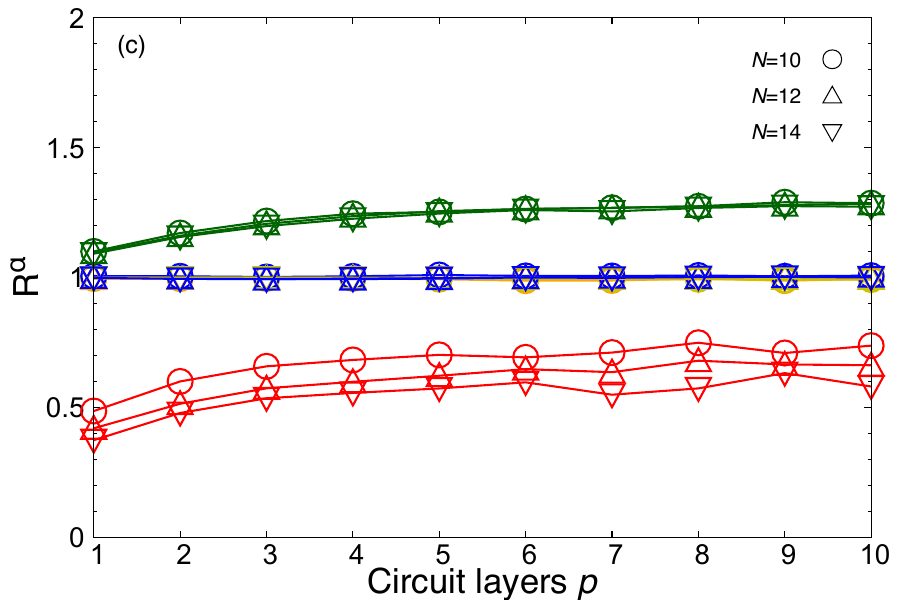}
    \includegraphics[width=0.45\linewidth]{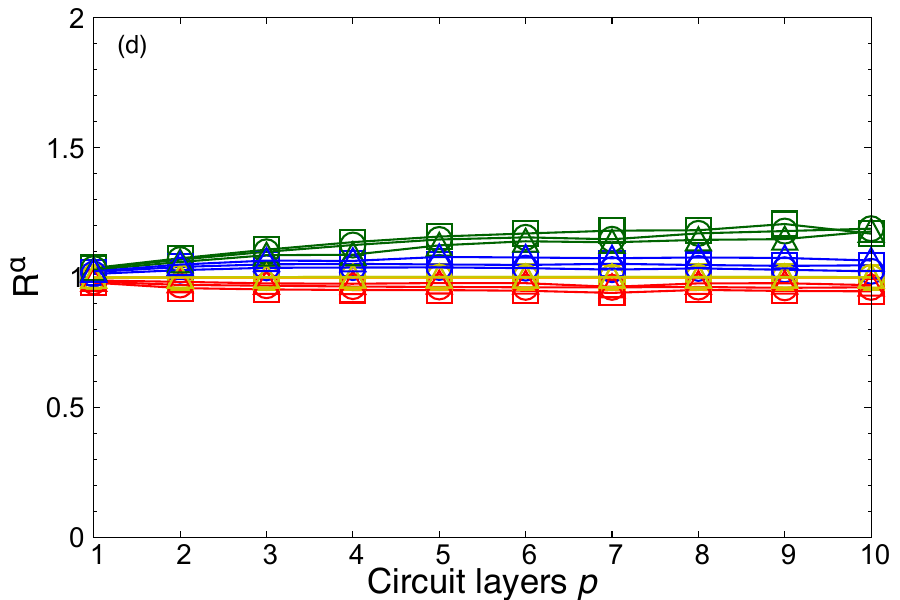}
    \caption{
    Dependences of $R^\alpha$ for $\alpha \in \{\mathrm{rpa}, \mathrm{shell}, \mathrm{g1}, \mathrm{g2}, \mathrm{cw} \}$ on the annealing time $\tau$ and the QAOA layers $p$ for (a),(c) the random Ising model and (b),(d) the REM.
    The approximation type is specified by color.
    The data for $R^\mathrm{shell}$ and $R^\mathrm{g1}$ are almost overlapped with $R^\mathrm{cw}$.
    Squares, circles, up-triangles, and down-triangles represent $N=8$, $10$, $12$, and $14$, respectively.
    }
    \label{fig:ratio_dE}
\end{figure}

To quantify the accuracy of each approximation, we define
\begin{align}
    R^\mathrm{rpa}=&\frac{d_E}{d_E^\mathrm{rpa}}, \quad  R^\mathrm{shell}=\frac{d_E^\mathrm{rpa}}{d_E^\mathrm{shell}}, \quad R^\mathrm{g1}=\frac{d_E^\mathrm{shell}}{d_E^\mathrm{g1}}, \nonumber\\R^\mathrm{g2}=&\frac{d_E^\mathrm{g1}}{d_E^\mathrm{g2}}, \quad R^\mathrm{cw}=\frac{d_E^\mathrm{g2}}{d_E^\mathrm{cw}}.
\end{align}
Thus, $R^\alpha \simeq 1$ indicates that the corresponding approximation introduces only a small quantitative error.
Figures~\ref{fig:ratio_dE} show $R^\alpha$ for $\alpha \in \{\mathrm{rpa}, \mathrm{shell}, \mathrm{g1}, \mathrm{g2}, \mathrm{cw} \}$ for the random Ising model and the REM.

For the random Ising model, the shell-uniformity, g1, and cw approximations are quantitatively accurate for both QA and QAOA, with the corresponding ratios remaining close to unity throughout the range of $\tau$ and $p$.
By contrast, the random-phase approximation shows a substantial deviation at small $\tau$ and $p$, while its accuracy improves as $\tau$ or $p$ increases.
$R^\mathrm{rpa}$ scales approximately as $N^{-1}$ for small $p$ and $\tau$.
This behavior is consistent with the increasing delocalization of the transition in Hamming space discussed in the main text.
The g2 approximation exhibits the opposite trend: its deviation from unity gradually increases with $\tau$ or $p$, reaching approximately $30$--$40\%$ in the largest-$\tau$ and largest-$p$ regimes considered here.
In contrast to the random-phase approximation, the deviation associated with the g2 approximation shows no appreciable  system-size dependence.

For the REM, the random-phase approximation is more accurate than for the random Ising model and shows a weak dependence on $\tau$ or $p$.
The shell-uniformity, g1, and cw approximations remain quantitatively accurate for both QA and QAOA.
As in the random Ising model, the deviation associated with the g2 approximation increases with $\tau$ or $p$.
No appreciable system-size dependence is observed for any of the approximations.

These results show that the qualitative difference between the random Ising model and the REM originates primarily from the random-phase approximation, whereas the subsequent approximations behave similarly in the two models.

Finally, we derive Eq.~(\ref{eq:var[w]}).
The covariance of $C_z$ and $w_z$ can be expressed as
\begin{align}
    \operatorname{Cov}[C,w]&=\frac{1}{2^N}\operatorname{Tr}[\hat C_{\mathrm{eff}}\hat W]
-\frac{1}{2^{2N}}\operatorname{Tr}[\hat C_{\mathrm{eff}}]\operatorname{Tr}[\hat W] \nonumber\\
    &=-\frac{1}{2^{N+1}} \sum_{i=1}^N \operatorname{Tr} [\hat{C}_\mathrm{eff} \hat{\sigma}_i^x] \nonumber\\
    &=-\frac{1}{2} \sum_{i=1}^N c_{\hat{X}_{\{i\}}}.
\end{align}
Here, we used the orthogonality of Pauli strings, $\operatorname{Tr}[\hat{P}_{\boldsymbol\mu}\hat{P}_{\boldsymbol\nu}]=2^N\delta_{\boldsymbol\mu,\boldsymbol\nu}$.
From the definition of the correlation coefficient,
\begin{equation}
    \operatorname{Var}[w]=(r^\mathrm{cw})^{-2} \frac{(\operatorname{Cov}[C,w])^2}{\operatorname{Var}[C]},
\end{equation}
which yields Eq.~(\ref{eq:var[w]}).

\section{Frustrated ring model}\label{appendix:1stQPT}
\begin{figure}[t!]
    \centering
    \includegraphics[width=0.9\linewidth]{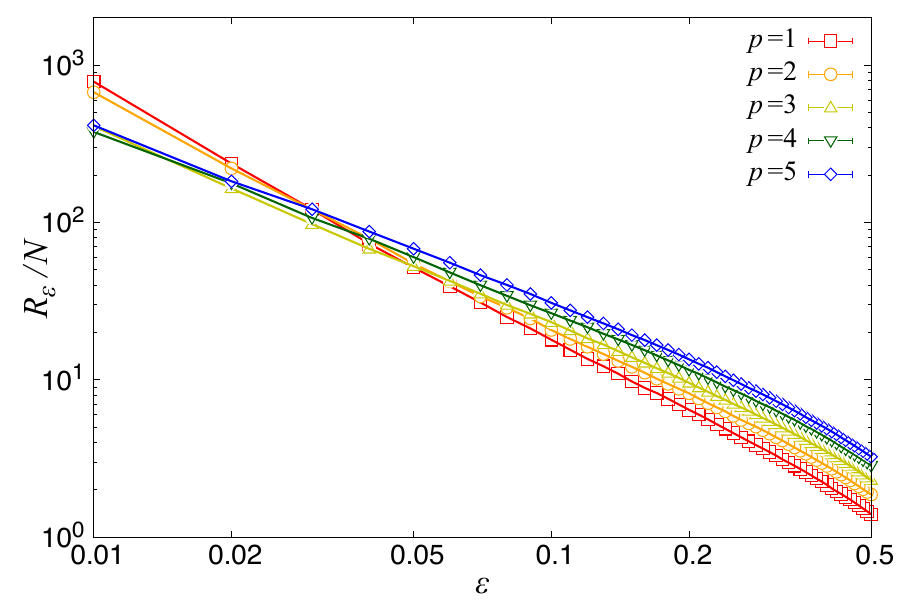}
    \caption{
    Quantum-echo local optimization results for the frustrated ring model: $R_\epsilon$ for different values of the number of layers $p$.
    We set $N=10001$.
    }
    \label{fig:Frustrated_Ring}
\end{figure}

We consider the frustrated ring model~\cite{cote2023diabatic} to investigate whether the performance of quantum-echo local optimization is affected by an exponentially small gap along the annealing path.

The cost function is given by
\begin{equation}
    C_z=-\sum_{i=1}^N J_i (-1)^{z_i + z_{i+1}},
\end{equation}
with periodic boundary condition (i.e., $z_{N+1}=z_1$) and odd $N$.
The model has three types of couplings:
\begin{equation}
    J_i=\left\{
    \begin{aligned}
        &-0.45   \text{ if } i=N\\
        &0.5    \text{ if } i=\frac{N\pm 1}{2}\\
        &1  \text{ otherwise}.
    \end{aligned}
    \right.
\end{equation}

We employ QAOA-based quantum-echo local optimization.
The transition matrix is essentially identical to that of the ferromagnetic Ising chain in the large $N$ limit.
The reason is that the two models differ only around a finite number of sites, while a finite-depth QAOA circuit probes only local structures within a finite distance.
Therefore, we use the same optimized variational parameters as those obtained for the ferromagnetic Ising chain.

Nevertheless, this argument does not immediately extend to the optimization performance.
Unlike the transition matrix, $R_\epsilon$ is an $O(N)$ dynamical quantity accumulated over many local updates, and therefore local disorder could influence long-time search dynamics.
Figure~\ref{fig:Frustrated_Ring} shows that $R_\epsilon$ qualitatively exhibits the same behavior as in the ferromagnetic Ising chain.
$R_\epsilon$ scales linearly with $N$, and large $p$ performs better for high-quality target configurations.
These results suggest that the optimization performance measured by $R_\epsilon$ is insensitive to the presence of an exponentially small energy gap along the annealing path.

\section{Numerical estimation of asymptotic amplitude $A_0$}\label{appendix:extrapolation}

\begin{table}[t]
\caption{
Comparison of the extrapolated amplitudes $A_{0,N}^{\mathrm{ext}}$ for
$N=10^5$ and $N=10^6$ with the theoretical estimates
$A_0^{\mathrm{diff}}=1/(2\pi D)$ obtained from the diffusion coefficients.
}
\label{tab:A0}
\begin{ruledtabular}
\begin{tabular}{cccc}
$p$ &
$10^3 A_{0,N}^\mathrm{ext}$ ($N=10^5$) &
$10^3 A_{0,N}^\mathrm{ext}$ ($N=10^6$) &
$A_0^{\mathrm{diff}} \times 10^3$ \\
\hline
1 & $74.5 \pm 7.2$ & $70.9 \pm 3.1$ & $70.8 \pm 1.0$ \\
2 & $31.5 \pm 4.2$ & $28.2 \pm 2.0$ & $27.8 \pm 0.4$ \\
3 & $13.0 \pm 1.2$ & $13.0 \pm 0.9$ & $12.3 \pm 0.2$ \\
4 & $8.1 \pm 2.6$ & $7.7 \pm 1.2$ & $6.6 \pm 0.1$ \\
5 & $5.1 \pm 1.6$ & $4.8 \pm 1.4$ & $3.9 \pm 0.1$ \\
\end{tabular}
\end{ruledtabular}
\end{table}

\begin{figure}[t!]
    \centering
    \includegraphics[width=0.9\linewidth]{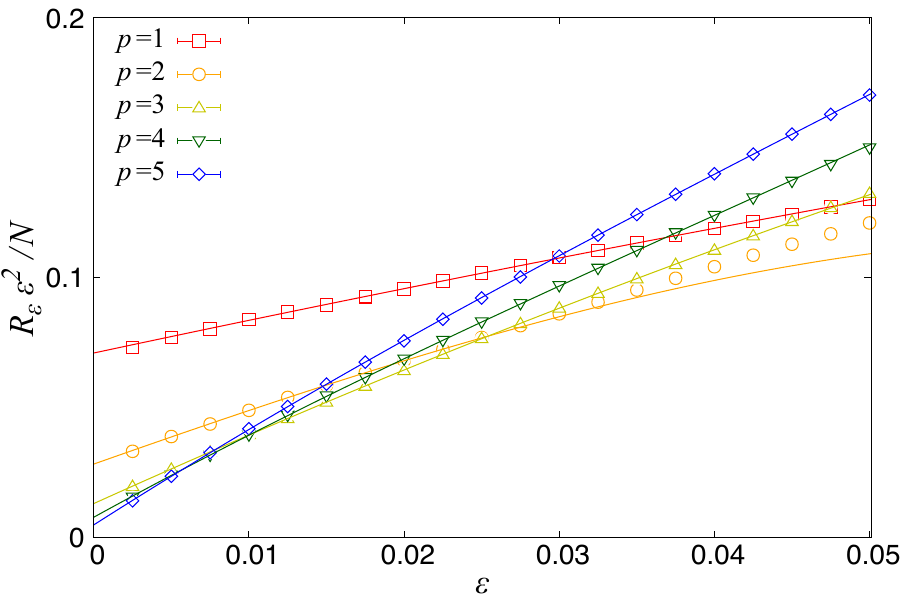}
    \caption{
    Dependences of $R_\epsilon \epsilon^2/N$ on $\epsilon$ for different values of $p$ at $N=10^6$.
    The curves show the polynomial fits whose extrapolated intervals are closest to the median of the intervals obtained using different fitting windows $[\epsilon_\mathrm{min},\epsilon_\mathrm{max}]$.
    }
    \label{fig:asymptotic_amplitude}
\end{figure}

We provide details of the numerical estimation of the asymptotic amplitude $A_0$ for each $p$.

First, based on the scaling $R_\epsilon \sim N \epsilon^{-2}$ in the diffusion-annihilation regime, we extrapolate $R_\epsilon \epsilon^2/N$ to $\epsilon=0$ to estimate $A_{0,N}^\mathrm{ext}$.
We fit the data at system size $N$ using quadratic or cubic polynomials over the interval $[\epsilon_\mathrm{min}, \epsilon_\mathrm{max}]$, where $\epsilon_\mathrm{min} \in \{0.0025, 0.005, 0.0075\}$ and $\epsilon_\mathrm{max} \in \{0.02, 0.03, 0.04, 0.05\}$.
We retain fits for which the reduced $\chi^2$ is less than unity and the number of residual degrees of freedom is at least three.

Table~\ref{tab:A0} presents the estimated values of $A_{0,N}^\mathrm{ext}$ and their uncertainties for different values of $p$ at $N=10^5$ and $10^6$.
We define $A_{0,N}^\mathrm{ext}$ as the median of the fitted intercepts, $A_{0,N}^\mathrm{ext}=\operatorname{Median}_j A^{(j)}$, where $A^{(j)}$ denotes the intercept obtained from the $j$th fit.
The uncertainty is estimated as $\delta A_{0,N}^\mathrm{ext}=\sqrt{\delta A_\mathrm{stat}^2 + \delta A_\mathrm{window}^2}$, where $\delta A_\mathrm{stat}$ is the median of the standard errors of the fitted intercepts and $\delta A_\mathrm{window}=\operatorname{max}_j|A^{(j)}-A_{0,N}^\mathrm{ext}|$ characterizes the sensitivity to the fitting procedure.
The uncertainty is dominated by $\delta A_\mathrm{window}$.
Figure~\ref{fig:asymptotic_amplitude} shows the dependence of $R_\epsilon \epsilon^2/N$ on $\epsilon$ for different values of $p$ at $N=10^6$, together with representative polynomial fits whose intercepts are closest to the median.

Second, we estimate $A_0$ using diffusion-annihilation theory~\cite{spouge1988exact}.
We denote this estimate by $A_0^\mathrm{diff}$.
In the long-time limit, the density of domain walls behaves as
\begin{equation}
    n_\mathrm{d} \sim \sqrt{\frac{1}{8\pi D t}}
\end{equation}
where $t$ is the reduced time, defined as the iteration count divided by the system size, and $D$ is the diffusion coefficient.
Using $\epsilon=2n_\mathrm{d}$ and $t=R_\epsilon/N$, we obtain $R_{\epsilon} \sim N/(2\pi D \epsilon^2)$, yielding
\begin{equation}
    A_0^\mathrm{diff}=\frac{1}{2\pi D}.
\end{equation}
We estimate the diffusion coefficient as
\begin{equation}
    D=\frac{1}{N_\mathrm{T}}\sum_{i=1}^{N_\mathrm{T}}\frac{\Delta x_i(T)^2}{2T},
\end{equation}
where $\Delta x_i(t)$ denotes the displacement of a domain wall from its initial position at time $t$ in the $i$th trial, and $N_\mathrm{T}$ is the number of trials.
We set $N=10^4$, $T=10^3$, and $N_\mathrm{T}=10$, starting from a configuration containing two domain walls at $x=0$ and $x=N/2$.
The observation time $T$ is chosen to be long enough for diffusive behavior to emerge but short enough to neglect interactions between the two domain walls.
We confirm that the estimated diffusion coefficient is nearly converged at $T=10^3$.

Table~\ref{tab:A0} shows that $A_{0,N}^\mathrm{ext}$ and $A_0^\mathrm{diff}$ agree within the estimated uncertainties.
Although $A_{0,N}^\mathrm{ext}$ at $N=10^6$ is slightly larger than $A_0^\mathrm{diff}$ for $p=4$ and $5$, the decrease in $A_{0,N}^\mathrm{ext}$ with increasing $N$ suggests that these deviations may diminish at larger system sizes.
Finally, we note that $A_{0,N}^{\mathrm{ext}}$ is estimated from the mean first-passage time $R_\epsilon$, whereas $A_0^{\mathrm{diff}}$ is derived from the mean domain-wall density $n_\mathrm{d}$.
The agreement between these estimates suggests that the distinction between the two quantities become less significant at large system sizes.

\bibliography{qemp}

\end{document}